\documentclass[final,5p,times,twocolumn]{elsarticle}
\graphicspath{ {./figures/} }
\usepackage{hyperref}
\usepackage{float}
\usepackage{verbatim} 
\usepackage{apalike}
\restylefloat{figure}
\usepackage{placeins}
\usepackage{xcolor}
\floatstyle{plaintop} 
\restylefloat{table}
\usepackage{amsmath,amssymb,amsfonts}
\usepackage{color,array}
\usepackage{booktabs}
\usepackage{multirow}
\usepackage{tabularx}
\usepackage{subcaption}
\usepackage{bbding}
\usepackage[utf8]{inputenc}
\usepackage[T1]{fontenc}
\usepackage{graphicx}

\journal{Expert Systems with Applications}

\biboptions{authoryear}

\begin{document}

\title{Enhancing Multimodal Medical Image Classification through Cross-Graph Modal Contrastive Learning}

\author[label1]{Jun-En Ding}
\ead{jding17@stevens.edu}

\author[label2]{Chien-Chin Hsu}
\ead{cchsu128@gmail.com}

\author[label3]{Chi-Hsiang Chu}
\ead{chchu1013@nuk.edu.tw}

\author[label4]{Shuqiang Wang}
\ead{sq.wang@siat.ac.cn}

\author[label1]{Feng Liu \corref{cor1}}
\ead{fliu22@stevens.edu}

\cortext[cor1]{Corresponding author.}

\address[label1]{Department of Systems Engineering, Stevens Institute of Technology, Hoboken, New Jersey, USA}
\address[label2]{Department of Nuclear Medicine, Kaohsiung Chang Gung Memorial Hospital, Kaohsiung, Taiwan}
\address[label3]{Institute of Statistics, National University of Kaohsiung, Kaohsiung, Taiwan}
\address[label4]{Shenzhen Institutes of Advanced Technology, Chinese Academy of Sciences, Shenzhen, China}

\begin{abstract}
The classification of medical images is a pivotal aspect of disease diagnosis, often enhanced by deep learning techniques. However, traditional approaches typically focus on unimodal medical image data, neglecting the integration of diverse non-image patient data. This paper proposes a novel Cross-Graph Modal Contrastive Learning (CGMCL) framework for multimodal structured data from different data domains to improve medical image classification. The model effectively integrates both image and non-image data by constructing cross-modality graphs and leveraging contrastive learning to align multimodal features in a shared latent space. An inter-modality feature scaling module further optimizes the representation learning process by reducing the gap between heterogeneous modalities. The proposed approach is evaluated on two datasets: a Parkinson's disease (PD) dataset and a public melanoma dataset. Results demonstrate that CGMCL outperforms conventional unimodal methods in accuracy, interpretability, and early disease prediction. Additionally, the method shows superior performance in multi-class melanoma classification. The CGMCL framework provides valuable insights into medical image classification while offering improved disease interpretability and predictive capabilities.  The code and data are available from \url{https://github.com/Ding1119/CGMCL}.
\end{abstract}


\begin{frontmatter}
\end{frontmatter}



\section{Introduction}
\label{introduction}

In recent years,  medical computing has increasingly shifted toward utilizing deep learning frameworks as a foundation for diagnostic analysis.  In particular, multimodal deep learning has emerged as a powerful approach for integrating diverse data sources to enhance predictive accuracy in various medical applications. The complex and heterogeneous nature of medical data, including radiological images, clinical records and histological images, necessitates the use of multimodal fusion methods to extract complementary features and improve clinical decision-making~\cite{yang2022large,nakach2024comprehensive,rupp2023exbehrt}. However, in real-world clinical settings, many modalities show inconsistency in modality distributions depending on disease characteristics, which indirectly creates limitations for multimodal learning~\cite{zhang2024multimodal}. Addressing these challenges, some studies have proposed advanced fusion architectures, such as feature-level fusion, decision-level fusion, and hybrid fusion, to optimize the integration of multimodal information~\cite{nakach2024comprehensive,tang2024joint}.  For example, integrating dermatological images with patient metadata significantly enhances the classification accuracy of skin cancer detection systems, outperforming unimodal image-based models~\cite{tang2024joint}.

Graph-based deep learning such as graph convolutional neural networks (GCNs)~\cite{kipf2017semi} learn from unstructured features by processing non-Euclidean distances, enabling them to perform various downstream tasks, including node classification, link prediction, and graph classification. An advanced framework of GCNs offer a promising avenue for identifying commonalities among patients in disease relationships via medical imaging data or symptom similarity~\cite{lu2021weighted,mao2022imagegcn}.  However, these methods only perform disease prediction on a single modality and lack information from both images and meta-features.  AMA-GCN \cite{chen2021ama} uses multi-layer aggregation mechanisms and automatic feature selection to integrate both image and non-image data for various diseases. Nevertheless, AMA-GCN's thresholds and aggregation methods require readjustment, and the model may overly depend on certain features, potentially leading to overfitting. Moreover, there is no unified approach for effectively achieving clinical interpretability through multimodal fusion, particularly for images and metadata associated with distinct graph structures in clinical settings, which should be further leveraged to enhance efficient latent space learning~\cite{wang2020gcn}. 

Recent studies have focused on aligning multimodal data in a shared embedding space for Alzheimer's disease (AD) prediction. Notably, attention mechanisms applied to tabular data have demonstrated improved model performance and interpretability~\cite{huang2023multimodal}, while multimodal transformers combining image and clinical data have shown success in predicting AD progression~\cite{liu2023triformer}. Furthermore, ~\cite{taleb2022contig} has demonstrated the effectiveness of self-supervised contrastive learning in handling different modalities through dedicated encoder networks, which map data into a shared feature space while optimizing a contrastive loss function. 

However, existing studies have primarily focused on specific diseases, with limited investigation into model generalization across diverse multimodal disease datasets and a lack of interpretable meta-features for disease classification. Recent approaches remain largely tailored to single-disease contexts and struggle to generalize across heterogeneous multimodal medical datasets that differ in data structure, modality distribution, and acquisition sources~\cite{teoh2024advancing,li2024review}. This heterogeneity, which arises from intrinsic disparities among medical images, quantitative biomarkers, and clinical metadata, presents substantial challenges for achieving robust multimodal fusion and interpretability~\cite{liu2025review}.


To address these limitations, we propose a novel  \underline{C}ross-\underline{G}raph \underline{M}odal \underline{C}ontrastive \underline{L}earning (CGMCL) framework that explicitly models multimodal heterogeneity through a dual-graph architecture. CGMCL constructs a cross-graph modal feature topology that represents diverse modalities, such as imaging data and quantitative parameters, as structured graphs. Using graph attention networks (GATs) \cite{Velickovic_2017}, our method learns node-level embeddings that capture inter-modal relationships, while a graph contrastive loss enforces cross-graph consistency and improves heterogeneous modality representation alignment. We validate the proposed CGMCL framework on two heterogeneous multimodal medical datasets, including a private PD dataset and a public melanoma dataset, to demonstrate its fusion capability and clinical interpretability. The main contributions of this study can be summarized as follows:


\begin{itemize}
    \item  We introduce a multimodal fusion framework that integrates multi-structured and heterogeneous modalities (e.g., medical images and meta-features) by representing patient characteristics as graph-structured features, thereby mitigating modality distribution inconsistencies and improving classification performance.

    \item To better evaluate the robustness of our proposed CGMCL model, we achieved more effective performance compared to fashion CNN-based models across two distinct types of multimodal datasets.
    
    \item  By incorporating meta-features into CGMCL, we enhance the model's interpretability for neurodegenerative disease diagnosis.

    \item We establish the first large-scale multimodal annotation dataset for early-stage PD patients and leverage CGMCL to quantify the relative importance of early degenerative symptoms, enabling better identification of early PD markers.
\end{itemize}

\section{RELATED WORK}

\subsection{SPECT for Parkinsonism Diagnosis}

\begin{figure}
\centering
\includegraphics[width=0.4\textwidth]{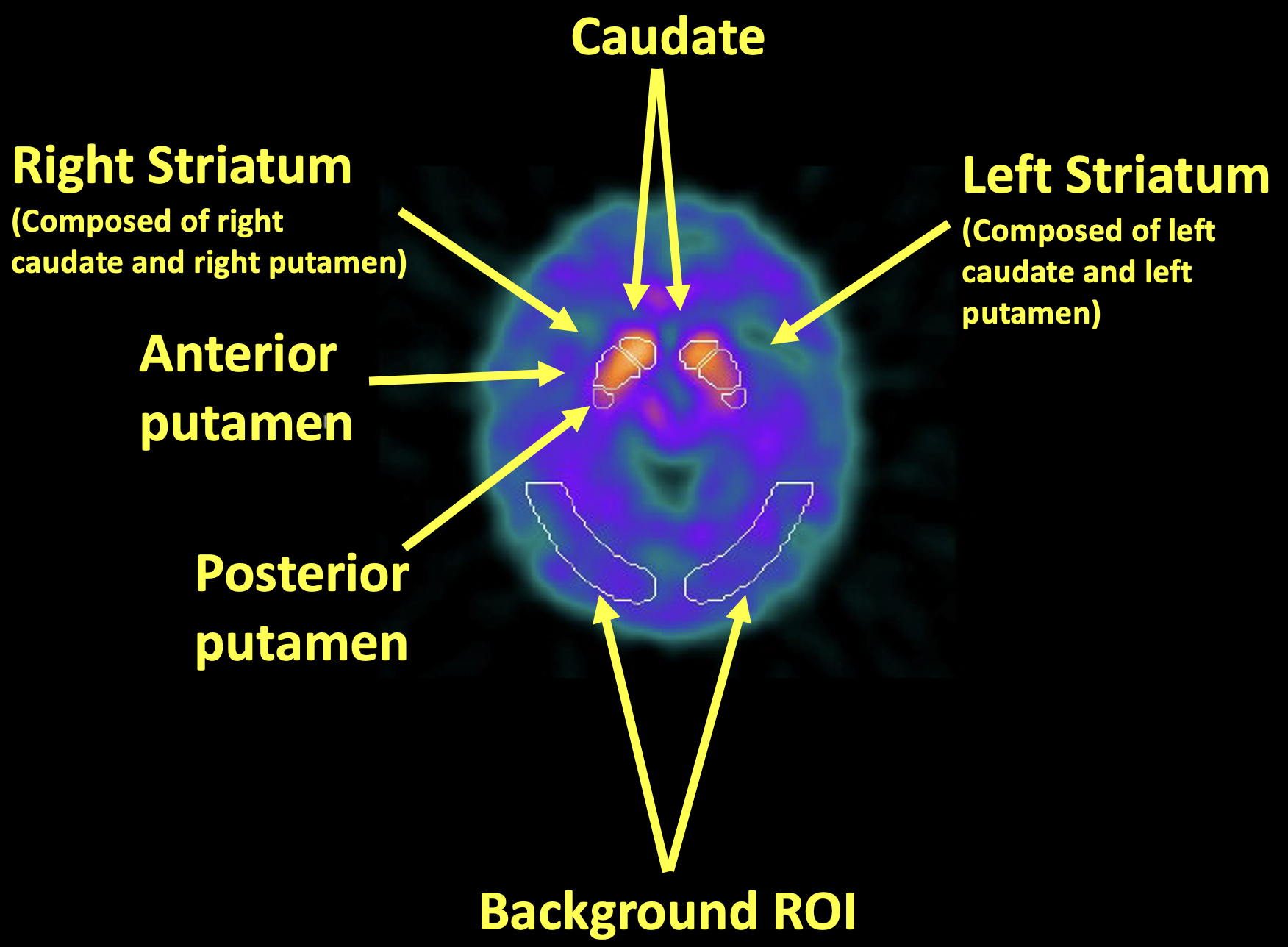}
\caption{The semi-quantification of parameters was derived from the DaTQUANT package and described the location of the striatum, relating to the entire striatum, caudate nucleus, entire putamen, anterior putamen, and posterior putamen~\cite{takatsu2023dysfunction}.}
\label{fig:semi-quantification}
\end{figure}

Dopamine imbalance generates toxic metabolites and oxidative stress, driving dopaminergic neuron degeneration and Parkinson's disease~\cite{bucher2020acquired,zhou2023role}. To assess the integrity of dopaminergic neurons, Dopamine transporter (DAT) imaging with single photon emission computed tomography (SPECT) can detect presynaptic dopaminergic neuronal dysfunction, serving as a valuable tool for differentiating PD from conditions without presynaptic dopaminergic deficit~\cite{brogley2019datquant}. For interpretation of DAT-SPECT, automatic semi-quantification of DAT-SPECT images with DaTQUANT package (GE Healthcare) provides objective parameters for assistance of image interpretation as shown in Fig.~\ref{fig:semi-quantification}.

Existing research has developed computer-aided diagnosis (CAD) systems for PD using DaTSCAN SPECT, which employ machine learning to analyze striatal dopamine transporter deficits. The approaches include high-dimensional voxel-based methods with dimensionality reduction~\cite{segovia2012improved,rojas2013application} and simplified striatal binding ratio (SBR) features that achieve accuracies exceeding $96$\%~\cite{prashanth2014automatic}. Among these, putamen SBR demonstrates superior discriminative power compared to caudate SBR, while logistic regression further enables risk probability estimation for patient stratification. However, single-modality SBR-based methods may not reflect the heterogeneous presentations encountered in routine clinical practice~\cite{serag2025multimodal,zhu2024multimodal}, particularly as high-dimensional voxel-based approaches require extensive feature reduction that potentially discards discriminative information~\cite{solana2021classification}.

\subsection{Multimodal Learning in Medical Imaging}

Traditional image-based models approaches often fail to capture the full complexity of diseases, whereas multimodal fusion leverages complementary information to enhance predictive performance~\cite{pahuja2022deep}. Some non-deep learning research works focus on using Wavelet Transform for multimodal image fusion in MRI/CT, PET/CT, and PET/MRI~\cite{zheng2007new,cheng2008medical,bhavana2015multi}. In neurodegenerative diseases diagnosis, studies have demonstrated that combining T1-weighted MRI, PET, and cerebrospinal fluid (CSF) biomarkers significantly improves early detection and progression monitoring of Parkinson’s Disease (PD) and Alzheimer’s Disease (AD)~\cite{dentamaro2024enhancing,pahuja2022deep}.

Recent research has introduced multimodal approaches for PD diagnosis and progression assessment. While traditional deep learning studies based on single modalities such as speech or imaging demonstrate useful discriminative power, they face limitations in generalizability~\cite{ul2022survey}. To address this limitation, ~\cite{lei2017joint} proposed a multimodal sparse learning framework that integrates magnetic resonance imaging (MRI) and diffusion tensor imaging (DTI) to simultaneously perform disease classification and clinical score prediction, demonstrating the robustness of multimodal fusion in PD diagnosis. Moreover, recent multimodal deep learning research focuses on PD classification applied multi-co-attention model successfully integrated brain single-photon emission computed tomography (SPECT) images with DNA methylation data~\cite{Taylor_2019}.  However, in the above studies, feature extraction from medical images has predominantly been conducted at the pixel level, with limited consideration for structured feature learning that accounts for inter-patient image similarities (e.g., graph structures). Therefore, we propose that early diagnosis, particularly for conditions such as PD, requires more precise multimodal models to improve early prediction and clinical interpretability.


\begin{figure}
\centering
\includegraphics[width=0.5\textwidth]{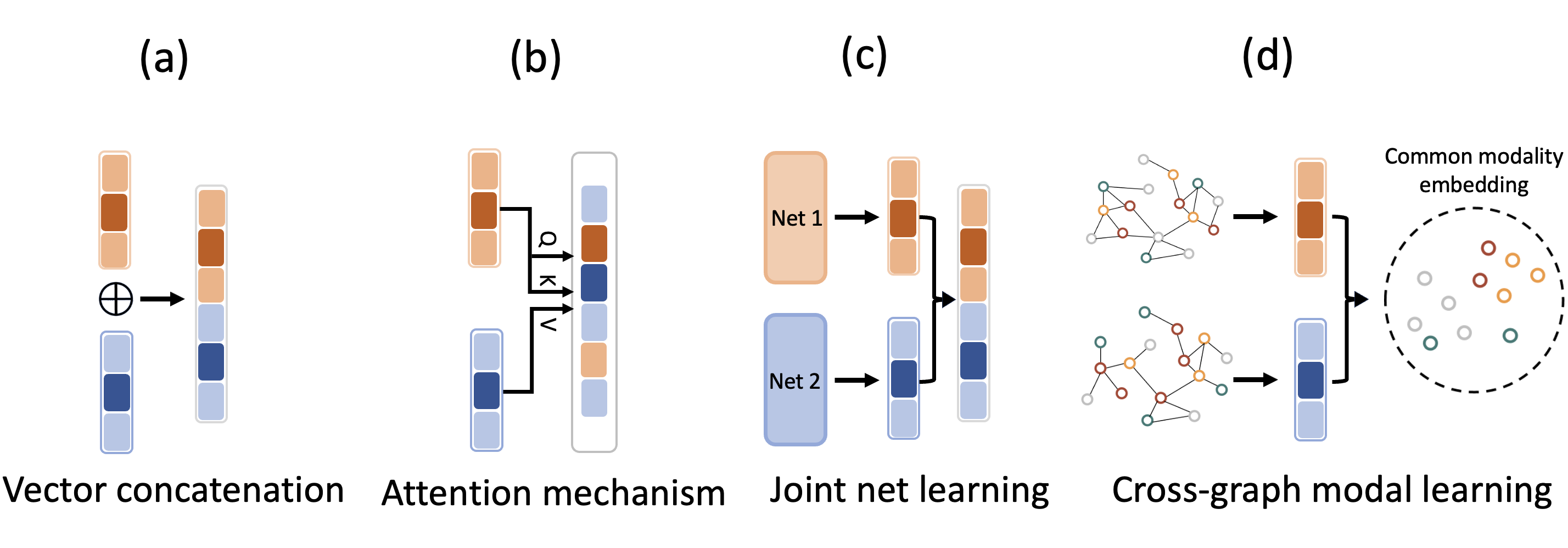}
\caption{The four neural network multimodal fusion methods are as follows: (a) and (b) represent conventional and widely-used vectors, with (a) utilizing vector concatenation and (b) employing attention-based modal learning. Method (c) uses a joint network for feature extraction from diverse modalities. Finally, method (d) illustrates our proposed cross-graph modal fusion, incorporating a graph structure.
}
\label{fig:fusion_method}
\end{figure}

\subsection{Multi-Modality Fusion Strategies}

According to deep learning fusion strategies, including feature extraction from two modalities, followed by fusion approaches such as vector concatenation, attention-based fusion, and joint network fusion learning to combine cross-domain features~\cite{cui2023deep,li2024review,duan2024deep,stahlschmidt2022multimodal,cui2023deep}, as shown in \textcolor{blue}{Fig.} \ref{fig:fusion_method} (a)-(c). Recent advancements in complicated multimodal framework have introduced more sophisticated fusion techniques, such as transformer-based fusion. MATR has demonstrated superior performance in multimodal medical image fusion by leveraging adaptive convolution and transformer mechanisms to utilize local and global contextual information~\cite{tang2022matr}. 

In particular, contrastive learning plays a crucial role in enhancing multimodal medical analysis under challenging conditions, such as limited data and class imbalance. For example, contrastive-based frameworks have been applied to few-shot diagnosis~\cite{yin2025cognitive}.  In medical image classification,~\cite{cong2024adaptive} developed AdUniGraph, an adaptive unified contrastive learning framework with graph-based feature aggregation, which employs an adaptive loss function and convolutional graph neural networks to effectively mitigate class imbalance issues in medical imaging datasets. Furthermore, \cite{cao2025multi} presented MCMCCL, a semi-supervised medical image segmentation model that integrates multi-view consistency learning and multi-scale cross-layer contrastive learning, enabling the more effective utilization of unlabeled data and enhancing segmentation accuracy. However, the greatest challenge with the approach mentioned above is that it does not consider the different characteristics of different modalities when performing modality fusion, which creates gaps in feature representation between their distributions.

\begin{figure*}
\centering
\includegraphics[width=1\textwidth]{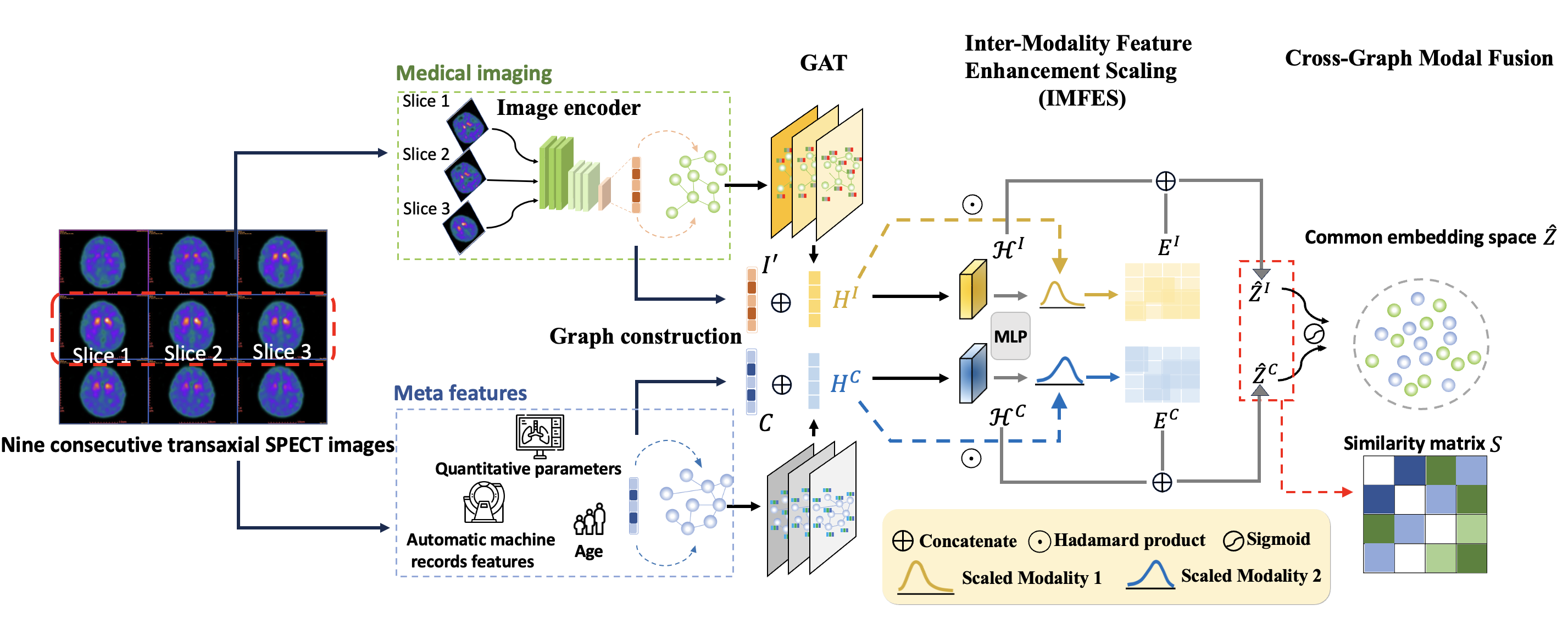}
\caption{The framework of multimodal cross-graph fusion for constructing a common feature space with contrastive learning.}

\label{fig:overall_framework}
\end{figure*}

To learn consistent latent representations, we designed a dual-graph architecture. The first graph utilizes patient image features extracted from a pre-trained ResNet model as node representations, where each node represents a patient. The second graph incorporates structured meta-features, as illustrated in \textcolor{blue}{Fig.} \ref{fig:fusion_method} (d). We then employ a dual-structured graph attention-based fusion framework to project and scale these different modalities into a consistent feature space, as shown in \textcolor{blue}{Fig.} \ref{fig:overall_framework}.



\section{METHODOLOGY}\label{sec2}

 The schematic overview of the proposed CGMCL framework be provided in Fig. \ref{fig:overall_framework}, illustrating how cross-modal representations are progressively aligned into a unified latent space. We propose dual-graph architecture that captures intra- and inter-modality dependencies through graph attention networks and a contrastive optimization objective to align heterogeneous features.

\subsection{Problem Definition and Formulation}

In clinical multimodal datasets, medical images and meta-features often have heterogeneous distributions and lack direct correspondence. To address this, we model patient similarities within each modality as separate graphs and align them through contrastive learning to improve disease classification beyond unimodal methods. Formally, given a multimodal set $\textbf{X}=\{X_{1}, X_{2},...,X_{N} \}$ for $N$ patients, each sample is defined as  3-tuple $X_{i} = \{(I_{i}, C_{i}, Y_{i}) \}_{i=1}^{N}$, where $I_{i} \in \mathbb{R}^{h \times h}$ denotes the medical image, $C_{i} \in \mathbb{R}^{F}$ the meta-features, and $Y_{i}$ corresponds to the disease labels. We first construct a non-linear model (e.g., an CNN) to generate initial feature maps $I_{i}^{\prime}$ from the images:

\begin{equation}\label{eq:eq1}
I_{i}^{\prime}=f(\cdot)=\mathrm{CNN}(I_{i}),
\end{equation}
where the function $f(\cdot): \mathbb{R}^{D} \to \mathbb{R}^{d}$ represents the feature extractor for images, $D$ is the dimension of the initial latent space in the $l-1$ layer, and $d$ is the reduced feature dimension after the $l$-th convolution layer. 

\subsection{Cross-Graph Representation Construction}

To overcome the limitations of conventional fusion methods, including feature concatenation and attention-based mechanisms, which tend to lose structural information among patients when heterogeneous modalities contain noise, we construct two similarity modality graphs, a structured-image graph $\mathcal{G}^{I}(\mathcal{E}^{I},\mathcal{V}^{I})$ and a structured meta-feature graph $\mathcal{G}^{C}(\mathcal{E}^{C},\mathcal{V}^{C})$, with edges $\left| \mathcal{E}^{I} \right|$ and $\left| \mathcal{E}^{C} \right|$ and vertices $\left| \mathcal{V}^{I} \right|$ and $\left| \mathcal{V}^{C} \right|$, respectively. Given the input encoded features $F_{i}$ (e.g., $I_{i}^{\prime}$ or $C_{i}$), we can construct the binary adjacency matrices $A^{I}$ and $A^{C}$ using a $K$-nearest neighbors graph~\cite{pedregosa2011scikit}.

\begin{equation}
A_{ij} =
\begin{cases}
1 &  \mbox{if }F_{i} \in \mathcal{N}(F_{j})  \hspace{8pt} or \hspace{8pt}F_{j} \in \mathcal{N}(F_{i})  \\
0 & \mbox{if } F_{i} \not\in \mathcal{N}(F_{j}) \hspace{8pt} or \hspace{5pt} F_{j} \not\in \mathcal{N}(F_{i}),
\end{cases}
\end{equation}
where $\mathcal{N}(\cdot)$ denotes the set of indices of the $K$ nearest neighbors of features $F_i$ and $F_j$ based on Euclidean distance. In a $K$-neighborhood, two data points $i$ and $j$ are connected by an edge $\mathcal{E}_{(i,j)}$ if $i$ is among the $K$ nearest neighbors of $j$, or vice versa. Each vertex $\mathcal{V}$ in the graph represents a patient.




\subsection{Graph Attention Encoder}

To encode both the image and clinical graphs, we adopt GAT that dynamically assigns weights to neighboring nodes by identifying the most influential patient connections during feature aggregation. Given the two input features node representations $I^{\prime} = [I^{\prime}_{1},I^{\prime}_{2},...,I^{\prime}_{N} ] \in \mathbb{R}^{N \times d}$ and $C = [C_{1},C_{2},...,C_{N} ] \in \mathbb{R}^{N \times F}$, we can formulate the multi-modality attention coefficients in a GAT as:

\begin{equation}
    e^{I^{\prime}}_{ij} = \sigma\left(\mathbf{W}^{I^{\prime}} I^{\prime}_i, \mathbf{W}^{I^{\prime}} I^{\prime}_j\right),
\end{equation}
\begin{equation}
    e^{C}_{ij} = \sigma\left(\mathbf{W}^{C} C_{i}, \mathbf{W}^{C} C_{j}\right),
\end{equation}
where $\sigma(\cdot)$  represents a non-linear transformation function (e.g., LeakyReLU, tanh), and the trainable weighted matrices are $\textbf{W}^{ I^{\prime}}$, $\textbf{W}^{C}$. We can then normalize the attention coefficients across neighboring nodes using the softmax function, which can be expressed as:

\begin{equation}
    \alpha^{I^{\prime}}_{ij} = \mathrm{softmax}_j (e^{I^{\prime}}_{ij}) = \frac{\exp(e^{I^{\prime}}_{ij})}{\sum_{k \in \mathcal{N}_i} \exp(e^{I^{\prime}}_{ik})},
\end{equation}
\begin{equation}
    \alpha^{C}_{ij} = \mathrm{softmax}_j (e^{C}_{ij}) = \frac{\exp(e^{C}_{ij})}{\sum_{k \in \mathcal{N}_i} \exp(e^{C}_{ik})},
\end{equation}
We apply weighted aggregation to the neighborhood node vectors using the normalized attention coefficients as attention scores:

\begin{equation}
\vspace{-1mm}
    H_i^{I^{\prime}} = \sigma \left( \sum_{j \in \mathcal{N}_i} \alpha^{I^{\prime}}_{ij} \mathbf{W}^{I^{\prime}} {I^{\prime}}_j \right),
\end{equation}
\begin{equation}
    H_i^C = \sigma \left( \sum_{j \in \mathcal{N}_i} \alpha^{C}_{ij} \mathbf{W}^C C_j \right),
\end{equation}
where $H^{I^{\prime}}$ and $H^{C}$ are the output representations from the GAT encoder.

We focus on enhancing the fusion of learned representations across cross-graph node modality embeddings. To integrate both the GAT output representations $H_i^{(I^{\prime})}$ and  $H_i^C$ with the encoder feature information, we also incorporate a concatenation of features generated by the CNN encoder $I^{\prime}$  and the meta-features $C$ into the graph encoder. This can be expressed as:

\begin{equation}\label{eq:eq9}
\mathcal{H}^{I} = \left[H_i^{(I')} \mathbin\Vert I_{i}^{\prime} \right],
\end{equation}
\begin{equation}\label{eq:eq10}
\mathcal{H}^{C}  = \left[H_i^C \mathbin\Vert C_{i} \right],
\end{equation}
where $\mathbin\Vert$ denotes the concatenation operator. The matrices $\mathcal{H}^{I} \in \mathbb{R}^{N \times (F+F^\prime)}$ and $\mathcal{H}^{C} \in \mathbb{R}^{N \times (F+F^\prime)}$ represent the concatenated feature representations. This concatenation jointly preserves the local graph topology and the intrinsic semantic information, thereby facilitating richer multimodal feature learning.

\subsection{Inter-Modality Feature Enhancement and Scaling}

To address scale discrepancies between features and reduce heterogeneity modality gaps in the latent space for two single-modality output features, we introduce the inter-modality feature enhancement and scaling (IMFES) module. This module is designed to enhance the learning of intrinsic modality distributions and preserve important structural information within each modality.
The process begins by applying a simple multilayer perceptron (MLP) to the concatenated tensors $\mathcal{H}^{I}$ and $\mathcal{H}^{C}$, transforming them into a single-modality probability matrix. Next, we perform element-wise multiplication between the GAT encoder outputs $H^{I^{\prime}}$ and $H^{C}$, and the non-linear MLP transforms $\zeta(\mathcal{H}^{I})$ and $\zeta(\mathcal{C}^{I})$, where $\zeta(\cdot) = \mathrm{MLP(\cdot)}$. This operation enables fine-grained extraction of individual modality features, as represented by the following equations:

\begin{equation}\label{eq:eq11}
E^{I} = H^{I^{\prime}} \odot \zeta(\mathcal{H}^{I}),
\end{equation}
\begin{equation}\label{eq:eq12}
E^{C} = H^{C} \odot \zeta(\mathcal{H}^{C}),
\end{equation}
where $E^{I}$  and $E^{C}$ are the resulting scaled feature matrices for the respective modalities, enhancing the representations learned by the GAT encoders, and $\odot$ denotes element-wise multiplication. Intuitively, IMFES can adaptively reweight to balance modality contributions and mitigate dominance from any single source.


\subsection{Cross-Graph Modal Fusion}


After feature refinement, the embeddings from both graphs are projected into a shared latent space through element-wise fusion. Specifically, each image-graph embedding $E^{I}$ and its corresponding graph representation $\mathcal{H}^{I}$ are first combined to form $\hat{Z}^{I} = [\mathcal{H}^{I} \mathbin\Vert E^{I}]$, while the clinical-graph embeddings $E^{C}$ and $\mathcal{H}^{C}$ are combined in the same way as $\hat{Z}^{C} = [\mathcal{H}^{C} \mathbin\Vert E^{C}]$. Then, the final image-graph embedding $\hat{Z}^{I}$ and the clinical-graph embedding $\hat{Z}^{C}$ are aligned by 
combining them via element-wise interaction and a nonlinear activation $\sigma(\cdot)$ to construct a unified latent space: $Z = \sigma(\hat{Z}^{I} + \hat{Z}^{C}).$ To further capture inter-patient relationships within this unified latent space, a similarity matrix $\mathbf{S} \in \mathbb{R}^{N \times N}$ is constructed to quantify pairwise correlations between patients:


\begin{equation}\label{eq:eq13}
S_{ij} =  Z_{i} \cdot (Z_{j})^{T}, \forall i,j \in \left [ 1,N \right ],
\end{equation}
where $S_{ij}$ encodes the joint similarity derived from both imaging and clinical modalities.

\subsection{Contrastive Loss Optimization} 



To enforce consistency across modalities, CGMCL introduces a cross-graph contrastive loss that simultaneously maximizes the similarity between positive patient pairs and increases the separation between dissimilar ones. The positive mask matrix is defined as $D_{pos} = \Theta(\hat{A}^{I} + \hat{A}^{C})$. In  contrast, the negative mask matrix is defined as $D_{neg} = \Theta\left((1 - \hat{A}^{I}) + (1 - \hat{A}^{C})\right)$, where $\hat{A}^{I}$ and $\hat{A}^{C}$ are adjacency matrices with self-loops. Here, $\Theta(\cdot)$ represents a threshold function, which can be expressed as:

\begin{equation}
\Theta(a) = \begin{cases}
1, & \text{if } a_{ij} \geq 0, \\
0, & \text{if } a_{ij} < 0,
\end{cases}
\end{equation}
where $a_{ij}$ represents elements in $\hat{A}^{I}$ or $\hat{A}^{C}$. 

We can calculate the positive and negative pairs in the similarity matrix as $S^+ = S \odot D_{pos}, \quad S^- = S \odot D_{neg}$. The sum of the positive and negative scores can then be calculate as:

\begin{equation}
     P_{s} = \sum_{i=1}^N \sum_{j=1}^N S_{ij}^+ Y_{j1},
\end{equation}
\begin{equation}
     N_{s} =  \sum_{i=1}^N \sum_{j=1}^N \left( \max\left(S_{ij}^- - \delta,\ 0\right) \right)^2 (1 - Y_{j1}),
\end{equation}
where $\delta > 0$ is the controllable margin, and $Y$ is the label matrix. The positive loss and negative loss can then be written as:

\begin{equation}\label{eq:eq17}
\mathcal{L}_{pos} = - \sum_{i=1}^N \log\left( P_{s} + \epsilon \right),
\end{equation}

\begin{equation}\label{eq:eq18}
\mathcal{L}_{neg} = - \sum_{i=1}^N \log\left( N_{s} + \epsilon \right),
\end{equation}
where $\epsilon $ is $1 \times 10^{-8}$  is used to prevent numerical computation issues. By using Eq.~\ref{eq:eq17} and~\ref{eq:eq18} we can ultimately obtain the combined losses, incorporating both the positive and negative loss, written as $\mathcal{L}_{contrastive} = \mathcal{L}_{pos} + \mathcal{L}_{neg}$. By minimizing $\mathcal{L}_{contrastive}$, the intra-class similarity is maximized, and the inter-class dissimilarity is increased.

For each modality processed within the graph-modal encoder, the modality-specific feature representations are independently projected into the label space via dedicated MLP classification heads. Formally, this process can be expressed as
\begin{equation}
\hat{y}_i^I = \text{MLP}_I(\hat{Z}^I_i),
\end{equation}

\begin{equation}
\hat{y}_i^C = \text{MLP}_C(\hat{Z}^C_i),
\end{equation}where $\hat{y}_i^I$ and $\hat{y}_i^C$ denote the predicted class probability vectors produced by the image–graph encoder and clinical–graph encoder, respectively.
Subsequently, for the final multi-modality prediction, the representations from each modality are combined and passed through a softmax activation as follows:
\begin{equation}
    \hat{y}^{*} = \arg\max\big(\text{softmax}(\hat{Z}^{I} + \hat{Z}^{C})\big).
\end{equation}
Here, $\hat{y}^{*}$ represents the final predicted class obtained by fusing complementary information from the image and clinical domains through a soft voting mechanism, where combined confidence scores from both modalities jointly determine the final decision and improve robustness.

To optimize the loss function and predict the probabilities of the final disease classes, we incorporated both $\hat{Z}^{I}$ and $\hat{Z}^{C}$ into the supervised binary classification loss using the softmax function. The cross-entropy loss function can be expressed as:

\begin{equation}\label{eq:I_loss}
\mathcal{L}_{I} = -\sum_{i=1}^{N}y_{i}^{T} \ln(\text{softmax}(\hat{y}_{i}^{I})),
\end{equation}


\begin{equation}\label{eq:C_loss}
\mathcal{L}_{C} = -\sum_{i=1}^{N}y_{i}^{T} \ln(\text{softmax}(\hat{y}_{i}^{C})),
\end{equation}
where $y_i$ is the one-hot vector of the true label, and $\hat{y}_i^I$ and $\hat{y}_i^C$ are the model's outputs for the image and meta-features, respectively. 

During the optimization process, we developed a comprehensive loss function that integrates cross-entropy and contrastive loss from the two cross-graph modalities. To further enhance the effectiveness of this combined loss, we incorporated the mean squared error between the similarity matrix $S$ and the diagonal matrix $D_{ii}=\sum_{i}A_{ii}$ when calculating the clustering of the module output fusion. The extend diagonal loss is expressed as:

\begin{equation}
\mathcal{L}_{diag} = \frac{1}{N}\sum^{N}_{i,j} (S_{ij} - D_{ii})^{2},
\end{equation}
where $\mathcal{L}_{diag}$ enforces structural consistency between graphs. We use $\beta$ as a leverage coefficient to control the optimization weight of the overall loss, which is defined as follows:
\begin{equation}\label{eq:eq22}
\mathcal{L}_{CGMCL} = (1- \beta)(\mathcal{L}_{I}  + \mathcal{L}_{C}) + \beta\mathcal{L}_{contrastive} + \mathcal{L}_{diag}, 
\end{equation}
where $\beta$ can be set between 0 and 1, and it controls the contribution level of different losses. The coefficient $\beta$ adjusts the weight assigned to each loss component. Specifically, $\mathcal{L}_I$ and $\mathcal{L}_C$ represent the supervised cross-entropy classification losses for the image modality and clinical meta-feature modality, respectively, as defined in Eq.~\ref{eq:I_loss} and Eq.~\ref{eq:C_loss}.

\begin{table}
\centering
\small
\setlength{\tabcolsep}{1pt} 
\caption{Summary statistics of demographics in PD patients with different subtypes}
\begin{tabular}{|l|c|c|c|}
\hline
\textbf{Subtypes} & \textbf{No. of Subjects}  & \textbf{Male(\%)} & \textbf{Age (Mean $\pm$ Std)} \\ \hline
Normal / Abnormal    &   127 / 154    &   54.6\%   & 67.5 $\pm$ 11.2      \\ \hline
Normal / MA    &    127 / 131    &    44.1\%    &    68.0 $\pm$ 12.0  \\ \hline
MA / Abnromal  &    131 / 154       &    55.8\%    &     68.4 $\pm$ 11.3    \\ \hline
\end{tabular}
\label{table:demographics}

\end{table}

\begin{figure}
\centering
\includegraphics[width=0.5\textwidth]{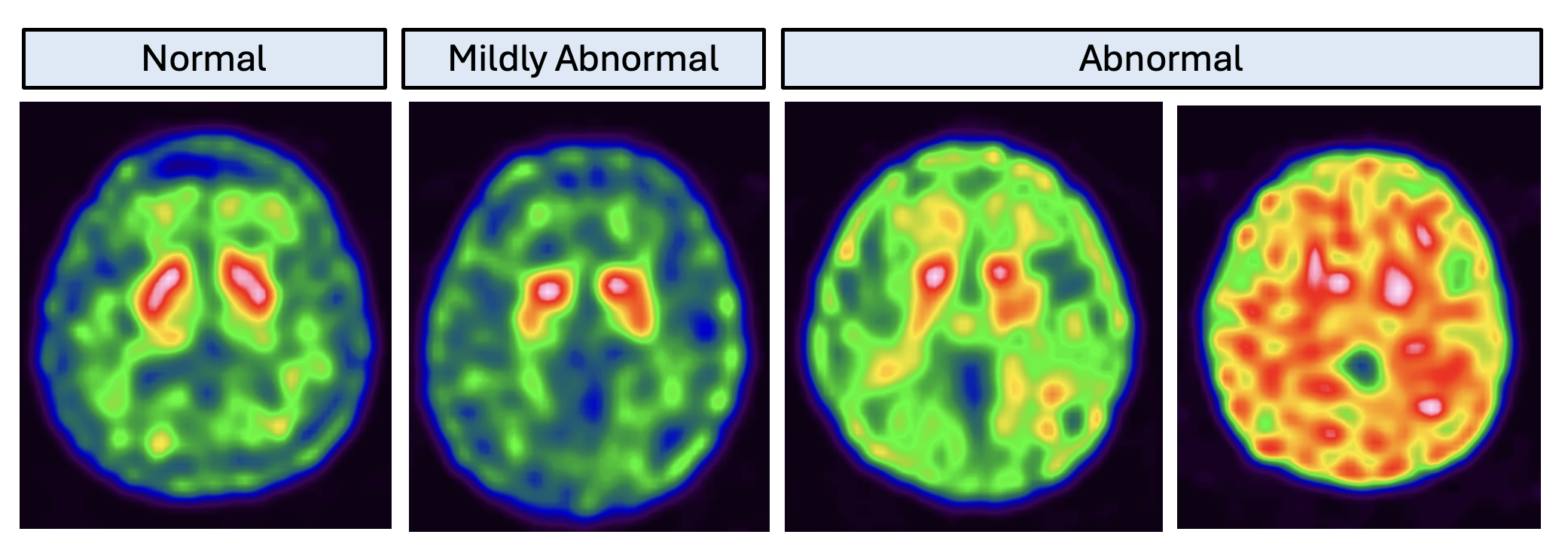}
\caption{Three subtypes of PD annotation. Early abnormalities typically manifest as unilateral putamen decline affecting the P/C ratio and symmetry, with progression involving AP and caudate until bilateral symmetric reduction; overall status is summarized by S = C + AP + PP.}

\label{fig:three_label}
\end{figure}

\section{Dataset Collection}

\begin{itemize}

    \item \textbf{Parkinson’s disease (PD)} Data for this study was collected at Kaohsiung Chang Gung Memorial Hospital, Taiwan, from January 2017 to June 2019, involving 416 patients~\cite{Ding_2021}. The study received approval from the Institutional Review Board, and all data were de-identified. Four expert nuclear medicine physicians annotated the data, labeling PD across three subtypes: Normal, Mildly Abnormal (MA), and Abnormal, as shown in Fig.~\ref{fig:three_label}. Table~\ref{table:demographics} and~\ref{table:clical_features_statistics} summarize the descriptive statistics of the demographics and meta-features.  
     
    The images used in this study were Tc99m TRODAT single-photon emission computed tomography (SPECT) scans acquired using a hybrid SPECT/CT system (Symbia T, Siemens Medical Solution). Image acquisition involved 30-second steps across 120 projections, covering a full 360-degree circular rotation with low-energy, high-resolution parallel-hole collimators. After reconstruction, CT-based attenuation correction imported the images into DaTQUANT for automatic semi-quantification of the DaTQUANT meta-features~\cite{Brogley_2019}. Twelve parameters were obtained from DaTQUANT: Striatum Right (S-R), Striatum Left (S-L), Anterior Putamen Right (AP-R), Anterior Putamen Left (AP-L), Posterior Putamen Right (PP-R), Posterior Putamen Left (PP-L), Caudate Right (C-R), Caudate Left (C-L), Putamen/Caudate Ratio Right (P/C-R), Putamen/Caudate Ratio Left (P/C-L), Putamen Asymmetry (PA), and Caudate Asymmetry (CA). The Fig.~\ref{fig:semi-quantification} contains additional details about these meta-features. The original SPECT images (800 × 1132) were resized to a standardized resolution of 128 × 128  pixels for model development.  A total of 412 preprocessed images and their twelve associated quantitative DaTQUANT meta-features were utilized for model training (n = 300) and testing (n = 112).

\begin{table}
\centering
\small
\caption{The mean ± standard deviation of twelve meta-features for three PD subtypes across different regions}
\begin{tabular}{lccc}
\toprule
\textbf{Meta-features} & \textbf{Normal} & \textbf{MA} & \textbf{Abnormal} \\
\midrule
S-R & 0.78 ± 0.22 & 0.65 ± 0.21 & 0.35 ± 0.16 \\
S-L & 0.78 ± 0.22 & 0.65 ± 0.21 & 0.36 ± 0.15 \\
AP-R & 0.81 ± 0.23 & 0.65 ± 0.23 & 0.32 ± 0.16 \\
AP-L & 0.83 ± 0.24 & 0.66 ± 0.22 & 0.35 ± 0.15 \\
PP-R & 0.60 ± 0.25 & 0.43 ± 0.22 & 0.23 ± 0.15 \\
PP-L & 0.58 ± 0.25 & 0.42 ± 0.22 & 0.23 ± 0.12 \\
C-R & 0.82 ± 0.23 & 0.73 ± 0.24 & 0.44 ± 0.21 \\
C-L & 0.83 ± 0.23 & 0.75 ± 0.24 & 0.46 ± 0.21 \\
P/C-R & 0.97 ± 0.08 & 0.93 ± 0.09 & 0.91 ± 0.09 \\
P/C-L & 0.96 ± 0.08 & 0.92 ± 0.09 & 0.91 ± 0.09 \\
PA & 0.04 ± 0.03 & 0.04 ± 0.04 & 0.06 ± 0.05 \\
CA & 0.04 ± 0.03 & 0.05 ± 0.04 & 0.06 ± 0.05 \\
\bottomrule
\end{tabular}
\label{table:clical_features_statistics}
\end{table}

    \item \textbf{Melanoma dataset \cite{kawahara2018seven}} The melanoma open dataset utilized for this study is a publicly available 7-point multimodal dataset comprising dermoscopic images and clinical data from 413 training and 395 testing samples. The classification task involves seven key image-based features: 1) pigment network (PN), 2) blue whitish veil (BWV), 3) vascular structures (VS), 4) pigmentation (PIG), 5) streaks (STR), 6) dots and globules (DaG), and 7) regression structures (RS). Additionally, the dataset includes five diagnostic categories: 1) basal cell carcinoma (BCC), 2) blue nevus (NEV), 3) melanoma (MEL), 4) miscellaneous (MISC), and 5) seborrheic keratosis (SK). The dermoscopic images have a resolution of 512 × 768 pixels, while the clinical data contains information on the patient's gender and lesion location. More melanoma categories and their annotation information can be referenced in Table 1 of the literature ~\cite{kawahara2018seven}

\begin{table}
\centering
\resizebox{\columnwidth}{!}{
\begin{tabular}{lccccc}
\hline
Model & Params(M) & GFLOPs & Latency(ms) & TrainTime(s) & GPU(MB) \\
\hline
ResNet18 & 11.178 & 67.880 & 8.522 & 31.99 & 1792.3 \\
\hline
Cluster-GT-T & 16.747 & 68.083 & 46.526 & 51.60 & 1859.5 \\
CoBFormer & 16.892 & 68.192 & 9.704 & 34.49 & 1960.0 \\
GTC & 18.233 & 68.952 & 9.787 & 34.12 & 2019.5 \\
GEAET & 18.728 & 68.315 & 9.460 & 33.77 & 2011.7 \\
GradFormer & 16.508 & 68.212 & 9.341 & 33.57 & \textbf{1970.9} \\
\hline
\textbf{CGMCL} & \textbf{15.070 ↑} & \textbf{68.052 ↑} & \textbf{8.879 ↑} & \textbf{32.94 ↑} & 1997.5 ↓ \\
\hline
\end{tabular}
}
\caption{Performance comparison of models sorted by training time and GPU usage in descending order. Arrows next to CGMCL indicate whether it outperforms (↑) or underperforms (↓) the best non-ResNet model in each metric.}
\label{tab:model_performance}
\end{table}

\end{itemize}

\begin{table*}[htbp]
\centering
\caption{The comparison of the proposed CGMCL's performance (mean $\pm$ std) between unimodal and multimodal approaches for PD subtypes: Normal vs MA. The best performance results are highlighted in \textbf{bold}, and "\_" indicates the second-best methods.}
\resizebox{\linewidth}{!}{%
{\Huge
\begin{tabular}{@{}lcccccccc@{}}
\toprule
& & & & \multicolumn{5}{c}{\textbf{Normal vs. MA}} \\ 
\cmidrule{5-9}
Models & Backbone & Image Feature Extractor & Clinical Features & ACC & SEN & SPE & PPV & NPV \\ 
\midrule
Logistic & - & - &  \CheckmarkBold & 0.63 $\pm$ 0.00 & 0.65 $\pm$ 0.01 & 0.62 $\pm$ 0.01 & 0.66 $\pm$ 0.02 & 0.61 $\pm$ 0.03 \\
XGboost & - & - &  \CheckmarkBold & 0.58 $\pm$ 0.00 & 0.56 $\pm$ 0.01 & 0.59 $\pm$ 0.02 & 0.61 $\pm$ 0.03 & 0.55 $\pm$ 0.03 \\
AdaBoost & - & - &  \CheckmarkBold & 0.56 $\pm$ 0.07 & 0.58 $\pm$ 0.12 & 0.55 $\pm$ 0.08 & 0.57 $\pm$ 0.08 & 0.56 $\pm$ 0.08 \\
\hline
\multicolumn{9}{l}{\textbf{Single slice (2-D)}} \\
\hline
 & - & 2-layer CNN & \XSolidBrush  & 0.65 $\pm$ 0.05 & 0.65 $\pm$ 0.06 & 0.64 $\pm$ 0.06 & 0.67 $\pm$ 0.06 & 0.62 $\pm$ 0.06 \\
Unimoal & - & VGG19 & \XSolidBrush & 0.71 $\pm$ 0.02 & 0.72 $\pm$ 0.03 & 0.70 $\pm$ 0.03 & 0.73 $\pm$ 0.03 & 0.69 $\pm$ 0.04 \\
& - & ResNet18 & \XSolidBrush & \underline{0.73 $\pm$ 0.05} & \underline{0.75 $\pm$ 0.06} & 0.69 $\pm$ 0.06 & 0.74 $\pm$ 0.05 & \underline{0.72 $\pm$ 0.06} \\
\hline
\multicolumn{9}{l}{\textbf{Three slices (3-D)}} \\
\hline
MHCA~\cite{taylor2019co}  & Transformer & 3D-CNN & \CheckmarkBold & 0.58 $\pm$ 0.02 & 0.52 $\pm$ 0.02 & 0.68 $\pm$ 0.03 & 0.68 $\pm$ 0.03 &  0.51 $\pm$ 0.02 \\
DeAF~\cite{li2023deaf} & Transformer & 3D-CNN & \CheckmarkBold & 0.62 $\pm$ 0.02 & 0.51 $\pm$ 0.11 & \underline{0.76 $\pm$ 0.10} & \underline{0.76 $\pm$ 0.06} & 0.54 $\pm$ 0.03 \\
TriFormer~\cite{liu2023triformer} & Transformer & 3D-CNN & \CheckmarkBold & 0.63 $\pm$ 0.02 & 0.56 $\pm$ 0.04 & 0.73 $\pm$ 0.02 & 0.74 $\pm$ 0.02 & 0.55 $\pm$ 0.02 \\
\hline
\multicolumn{9}{l}{\textbf{Graph-Transformer Models (2-D)}} \\
\hline
GEAET~\cite{GEAET_cite} & Graph-Transformer &  ResNet18 & \CheckmarkBold & 0.71 $\pm$ 0.03 & 0.70 $\pm$ 0.06 & 0.73 $\pm$ 0.06 & 0.74 $\pm$ 0.05 & 0.68 $\pm$ 0.05 \\
CoBFormer~\cite{Xing2024Less} & Grpah-Transformer &  & \CheckmarkBold & 0.72 ± 0.04 & 0.72 ± 0.06 &  0.71 ± 0.06  & 0.73 ± 0.04 &  0.70 ± 0.05 \\
GTC~\cite{sun2025gtc} & Mixture Graph-Transformer &  & \CheckmarkBold & 0.73 ± 0.04 & 0.69 ± 0.07 & 0.77 ± 0.06 & 0.78 ± 0.07 & 0.69 ± 0.05 \\
GradFormer~\cite{gradformer_cite} & Graph-Transformer &  & \CheckmarkBold & 0.72 ± 0.02 &  0.69 ± 0.06 & 0.76 ± 0.05 &  0.77 ± 0.05 & 0.69 ± 0.04 \\
Cluster-GT (N2C-Attn-T)~\cite{huang2024cluster} & Graph-Transformer &  & \CheckmarkBold & 0.72 ± 0.03 & 0.68 ± 0.07 & 0.77 ± 0.06 & 0.77 ± 0.06  &  0.68 ± 0.05 \\
Cluster-GT (N2C-Attn-L)~\cite{huang2024cluster} & Graph-Transformer &  & \CheckmarkBold & 0.74 ± 0.02 & 0.74 ± 0.03 & 0.75 ± 0.04   & 0.77 ± 0.04 & 0.71 ± 0.03 \\
\hline
\multicolumn{9}{l}{\textbf{Proposed Model (2-D)}} \\
\hline
CGMCL & GCN+GCN & ResNet18 & \CheckmarkBold & 0.70 $\pm$ 0.05 & 0.66 $\pm$ 0.06 & 0.75 $\pm$ 0.08 & 0.75 $\pm$ 0.08 & 0.66 $\pm$ 0.05 \\
CGMCL & GCN+GCN & VGG19 & \CheckmarkBold & 0.65 $\pm$ 0.03  & 0.65 $\pm$ 0.05 & 0.65 $\pm$ 0.02 & 0.67 $\pm$ 0.03 & 0.63 $\pm$ 0.04 \\
CGMCL & GAT+GAT & VGG19 & \CheckmarkBold & 0.70 $\pm$ 0.04  & 0.69 $\pm$ 0.06 & 0.69 $\pm$ 0.04 & 0.71 $\pm$ 0.05 & 0.66 $\pm$ 0.05 \\
CGMCL & GAT+GAT & ResNet18 & \CheckmarkBold & \textbf{0.76 $\pm$ 0.03}  & \textbf{0.75 $\pm$ 0.03} & \textbf{0.78 $\pm$ 0.05} & \textbf{0.79 $\pm$ 0.05} & \textbf{0.73 $\pm$ 0.04} \\
\hline
Comparison with \textbf{Unimodal} Improvements (\%) & & & - & +4.11\%$^{**}$ & +0.00\%$^{}$ & +13.04\%$^{***}$ & +6.76\%$^{***}$ & +1.39\%$^{}$ \\
\hline
Comparison with \textbf{Transformer-based} Improvements (\%) & & & - & +20.63\%$^{***}$ & +33.93\%$^{***}$ & +2.63\%$^{}$ & +3.95\%$^{*}$ & +32.73\%$^{***}$ \\
\hline
Comparison with \textbf{Graph-Transformer} Improvements (\%) & & & - & +2.70\%$^{**}$ & +1.35\%$^{}$ & +1.30\%$^{}$ & +1.28\%$^{}$ & +2.82\%$^{*}$ \\

\bottomrule
\end{tabular}%
}}
\vspace{1mm}
\footnotesize\\
\textit{Significance.} $^{*}$p $<$ 0.05, $^{**}$p $<$ 0.01, $^{***}$p $<$ 0.001 indicate statistically significant improvements of CGMCL compared to the corresponding baseline models.
\label{table:Normal_MA}
\end{table*}

\begin{table*}[htbp]
\centering
\caption{The comparison of the proposed CGMCL's performance (mean $\pm$ std) between unimodal and multimodal approaches for PD subtypes: Abnormal vs Abnormal. The best performance results are highlighted in \textbf{bold}, and "\_" indicates the second-best methods}
\resizebox{\linewidth}{!}{%
{\Huge
\begin{tabular}{@{}lcccccccc@{}}
\toprule
& & & & \multicolumn{5}{c}{\textbf{MA vs. Abnormal}} \\ 
\cmidrule{5-9}
Models & Backbone & Image Feature Extractor & Clinical Features & ACC & SEN & SPE & PPV & NPV \\ 
\midrule
Logistic & - & - & \CheckmarkBold & 0.86 $\pm$ 0.00 & \underline{0.88 $\pm$ 0.02} & 0.84 $\pm$ 0.02 & 0.85 $\pm$ 0.01 & \underline{0.87 $\pm$ 0.01} \\
XGboost & - & - & \CheckmarkBold & 0.83 $\pm$ 0.00 & 0.83 $\pm$ 0.00 & 0.82 $\pm$ 0.02 & 0.83 $\pm$ 0.02 & 0.82 $\pm$ 0.01 \\
AdaBoost & - & - & \CheckmarkBold & 0.75 $\pm$ 0.03  & 0.73 $\pm$ 0.09 & 0.79 $\pm$ 0.09 & 0.80 $\pm$ 0.09 & 0.71 $\pm$ 0.09 \\
\hline
\multicolumn{9}{l}{\textbf{Single slice (2-D)}} \\
\hline
& - & 2-layer CNN & \XSolidBrush  & 0.83 $\pm$ 0.04 & 0.84 $\pm$ 0.03 & 0.82 $\pm$ 0.05 & 0.83 $\pm$ 0.04 & 0.83 $\pm$ 0.03 \\
Unimoal & - & VGG19 & \XSolidBrush & 0.85 $\pm$ 0.03 & 0.85 $\pm$ 0.03 & 0.85 $\pm$ 0.04 & 0.85 $\pm$ 0.03 & 0.87 $\pm$ 0.03 \\
 & - & ResNet18 & \XSolidBrush & 0.87 $\pm$ 0.04 & 0.86 $\pm$ 0.04 & 0.86 $\pm$ 0.04 & 0.88 $\pm$ 0.05 & 0.87 $\pm$ 0.04 \\
\hline

\hline
\multicolumn{9}{l}{\textbf{Three slices (3-D)}} \\
\hline
MHCA~\cite{taylor2019co}  & Transformer & 3D-CNN & \CheckmarkBold & 0.80 $\pm$ 0.03 & 0.74 $\pm$ 0.04  & 0.88 $\pm$ 0.04 & 0.90 $\pm$ 0.03  &  0.71 $\pm$ 0.04 \\
DeAF~\cite{li2023deaf} & Transformer & 3D-CNN & \CheckmarkBold & 0.87 $\pm$ 0.03 & 0.84 $\pm$ 0.06 & 0.91 $\pm$ 0.03 & \textbf{0.93 $\pm$ 0.02} & 0.80 $\pm$ 0.06 \\
TriFormer~\cite{liu2023triformer} & Transformer & 3D-CNN & \CheckmarkBold & 0.83 $\pm$ 0.03 & 0.79 $\pm$ 0.03 & 0.89 $\pm$ 0.03 & 0.91 $\pm$ 0.02 & 0.75 $\pm$ 0.03 \\
\hline
\multicolumn{9}{l}{\textbf{Graph-Transformer Models (2-D)}} \\
\hline
GEAET~\cite{GEAET_cite} & Graph-Transformer & ResNet18 & \CheckmarkBold & 0.88 ± 0.02 & 0.88 ± 0.04 & 0.88 ± 0.03 & 0.88 ± 0.03 & 0.88 ± 0.03 \\
CoBFormer~\cite{Xing2024Less} & Graph-Transformer &  & \CheckmarkBold & 0.85 ± 0.01 & 0.84 ± 0.02 & 0.85 ± 0.03 & 0.85 ± 0.02 & 0.84 ± 0.01 \\
GTC~\cite{sun2025gtc} & Mixture Graph-Transformer &  & \CheckmarkBold & 0.87 ± 0.02 & 0.86 ± 0.02 & 0.87 ± 0.02 & 0.88 ± 0.02 & 0.86 ± 0.02 \\
GradTransformer~\cite{gradformer_cite} & Graph Transformer &  & \CheckmarkBold & 0.88 ± 0.01 & 0.87 ± 0.02 & 0.89 ± 0.02 & 0.89 ± 0.02 & 0.87 ± 0.01 \\
Cluster-GT (N2C-Attn-T)~\cite{huang2024cluster} & Graph-Transformer &  & \CheckmarkBold & 0.86 ± 0.01 & 0.84 ± 0.03 & 0.87 ± 0.04 & 0.87 ± 0.03 & 0.85 ± 0.02 \\
Cluster-GT (N2C-Attn-L)~\cite{huang2024cluster} & Graph-Transformer & - & \CheckmarkBold & 0.84 ± 0.02 & 0.85 ± 0.03 & 0.84 ± 0.01 & 0.84 ± 0.01 & 0.85 ± 0.03 \\
\hline
\multicolumn{9}{l}{\textbf{Proposed Model (2-D)}} \\
\hline
CGMCL& GCN+GCN & ResNet18 & \CheckmarkBold & \underline{0.87 $\pm$ 0.02} & 0.86 $\pm$ 0.04 & \underline{0.89 $\pm$ 0.03} & 0.89 $\pm$ 0.03 & 0.86 $\pm$ 0.03 \\
CGMCL & GCN+GCN & VGG19 & \CheckmarkBold & 0.84 $\pm$ 0.01  & 0.82 $\pm$ 0.02 & 0.85 $\pm$ 0.03 & 0.85 $\pm$ 0.02 & 0.82 $\pm$ 0.02 \\
CGMCL & GAT+GAT & VGG19 & \CheckmarkBold & 0.83 $\pm$ 0.04  & 0.83 $\pm$ 0.04 & 0.82 $\pm$ 0.04 & 0.83 $\pm$ 0.04 & 0.82 $\pm$ 0.04 \\
CGMCL & GAT+GAT & ResNet18 & \CheckmarkBold & \textbf{0.89 $\pm$ 0.01}  & \textbf{0.88 $\pm$ 0.02} & \textbf{0.91 $\pm$ 0.03} & \underline{0.91 $\pm$ 0.03} & \textbf{0.88 $\pm$ 0.02} \\

\hline

Comparison with \textbf{Unimodal} Improvements (\%) & & & - & +2.30\%$^{**}$ & +0.00\%$^{}$ & +5.81\%$^{***}$ & +3.41\%$^{**}$ & +1.15\%$^{}$ \\
\hline
Comparison with \textbf{Transformer-based} Improvements (\%) & & & - & +2.30\%$^{***}$ & +4.76\%$^{***}$ & +0.00\%$^{}$ & 0.00\%$^{}$ & +10.00\%$^{***}$ \\
\hline
Comparison with \textbf{Graph-Transformer} Improvements (\%) & & & - & +1.14\%$^{**}$ & +0.00\%$^{}$ & +2.25\%$^{**}$ & +2.25\%$^{**}$ & +0.00\%$^{}$ \\

\bottomrule
\end{tabular}%
}}
\vspace{1mm}
\footnotesize\\
\textit{Significance.} $^{*}$p $<$ 0.05, $^{**}$p $<$ 0.01, $^{***}$p $<$ 0.001 indicate statistically significant improvements of CGMCL compared to the corresponding baseline models.
\label{table:MA_Abnormal}
\end{table*}


\begin{table*}[htbp]
\centering
\caption{The comparison of the proposed CGMCL's performance (mean $\pm$ std) between unimodal and multimodal approaches for PD subtypes: Normal vs Abnormal. The best performance results are highlighted in \textbf{bold}, and "\_" indicates the second-best methods}

\resizebox{\linewidth}{!}{%
{\Huge
\begin{tabular}{@{}lcccccccc@{}}
\toprule
& & & & \multicolumn{5}{c}{\textbf{Normal vs. Abnormal}} \\
\cmidrule{5-9}
Models & Backbone & Image Feature Extractor & Clinical Features & ACC & SEN & SPE & PPV & NPV \\
\midrule
Logistic & - & - & \CheckmarkBold & 0.80 $\pm$ 0.00 & 0.80 $\pm$ 0.01 & 0.81 $\pm$ 0.01 & 0.80 $\pm$ 0.01 & 0.81 $\pm$ 0.01 \\
XGboost & - & - & \CheckmarkBold & 0.80 $\pm$ 0.00 & 0.77 $\pm$ 0.00 & 0.80 $\pm$ 0.02 & 0.78 $\pm$ 0.01 & 0.79 $\pm$ 0.03 \\
AdaBoost & - & - & \CheckmarkBold & 0.79 $\pm$ 0.04 & 0.78 $\pm$ 0.06 & 0.79 $\pm$ 0.05 & 0.78 $\pm$ 0.03 & 0.79 $\pm$ 0.06 \\
\hline
\multicolumn{9}{l}{\textbf{Single slice (2-D)}} \\
\hline
& - & 2-layer CNN & \XSolidBrush  & 0.86 $\pm$ 0.02 & 0.86 $\pm$ 0.03 & 0.87 $\pm$ 0.03 & 0.87 $\pm$ 0.03 & 0.86 $\pm$ 0.03 \\
Unimodal & - & VGG19 & \XSolidBrush & 0.87 $\pm$ 0.09 & 0.86 $\pm$ 0.01 & 0.88 $\pm$ 0.01 & 0.88 $\pm$ 0.02 & 0.86 $\pm$ 0.01 \\
& - & ResNet18 & \XSolidBrush & \underline{0.90 $\pm$ 0.02} & 0.89 $\pm$ 0.02 & \underline{0.91 $\pm$ 0.03} & \underline{0.91 $\pm$ 0.03} & 0.89 $\pm$ 0.03 \\

\hline
\multicolumn{9}{l}{\textbf{Three slices (3-D)}} \\
\hline
MHCA~\cite{taylor2019co} & Transformer & 3D-CNN & \CheckmarkBold & 0.87 $\pm$ 0.01 & \textbf{0.92 $\pm$ 0.03} & 0.82 $\pm$ 0.02 & 0.85 $\pm$ 0.01 &  0.90 $\pm$ 0.03 \\
DeAF~\cite{li2023deaf} & Transformer & 3D-CNN & \CheckmarkBold & 0.88 $\pm$ 0.01 & 0.87 $\pm$ 0.02 & 0.89 $\pm$ 0.01 & 0.90 $\pm$ 0.00 & 0.86 $\pm$ 0.02 \\
TriFormer~\cite{liu2023triformer} & Transformer  & 3D-CNN & \CheckmarkBold & 0.83 $\pm$ 0.03 & 0.86 $\pm$ 0.03 & 0.80 $\pm$ 0.03 & 0.83 $\pm$ 0.03 & 0.84 $\pm$ 0.03 \\

\hline

\multicolumn{9}{l}{\textbf{Graph-Transformer Models (2-D)}} \\
\hline
GEAET~\cite{GEAET_cite} & Graph-Transformer & ResNet18 & \CheckmarkBold & 0.90 $\pm$ 0.01 & 0.89 $\pm$ 0.02 & 0.91 $\pm$ 0.02 & 0.91 $\pm$ 0.02 & 0.89 $\pm$  0.02 \\
CoBFormer~\cite{Xing2024Less} & Graph-Transformer &  & \CheckmarkBold & 0.90 ± 0.01 & 0.89 ± 0.01 & 0.90 ± 0.02 & 0.90 ± 0.02 & 0.89 ± 0.01 \\
GTC~\cite{sun2025gtc} & Mixture Graph-transformer &  & \CheckmarkBold & 0.89 ± 0.01 & 0.88 ± 0.02 & 0.90 ± 0.02 & 0.91 ± 0.02 & 0.88 ± 0.02 \\
GradFormer~\cite{gradformer_cite} & Graph-Transformer &  & \CheckmarkBold & 0.88 ± 0.01 & 0.87 ± 0.02 & 0.89 ± 0.02 & 0.89 ± 0.02 & 0.87 ± 0.02  \\
Cluster-GT-T~\cite{huang2024cluster} & Graph-Transformer &  & \CheckmarkBold & 0.88 ± 0.01 & 0.86 ± 0.03 & 0.91 ± 0.04 & 0.91 ± 0.03 & 0.86 ± 0.02 \\
Cluster-GT-L~\cite{huang2024cluster} & Graph-Transformer &  & \CheckmarkBold & 0.87 ± 0.02 & 0.86 ± 0.04 & 0.88 ± 0.02 & 0.89 ± 0.03 & 0.86 ± 0.03  \\

\hline
\multicolumn{9}{l}{\textbf{Proposed Model (2-D)}} \\
\hline
CGMCL& GCN+GCN & ResNet18 & \CheckmarkBold & 0.87 $\pm$ 0.01 & 0.86 $\pm$ 0.02 & 0.89 $\pm$ 0.02 & 0.89 $\pm$ 0.02 & 0.86 $\pm$ 0.02 \\
CGMCL & GCN+GCN & VGG19 & \CheckmarkBold & 0.88 $\pm$ 0.01 & 0.87 $\pm$ 0.02 & 0.89 $\pm$ 0.02 & 0.89 $\pm$ 0.02 & 0.87 $\pm$ 0.02 \\
CGMCL & GAT+GAT & VGG19 & \CheckmarkBold & 0.89 $\pm$ 0.01 & 0.89 $\pm$ 0.02 & 0.89 $\pm$ 0.01 & 0.90 $\pm$ 0.01 & \underline{0.89$\pm$ 0.02} \\
CGMCL & GAT+GAT & ResNet18 & \CheckmarkBold & \textbf{0.90 $\pm$ 0.01} & \underline{0.90 $\pm$ 0.02} & \textbf{0.91 $\pm$ 0.02} & \textbf{0.91 $\pm$ 0.02} & \textbf{0.89 $\pm$ 0.02} \\
\hline

Comparison with \textbf{Unimodal} Improvements (\%) & & & - & +0.00\%$^{}$ & +1.12\%$^{*}$ & +0.00\%$^{}$ & +0.00\%$^{}$ & +0.00\%$^{}$ \\
\hline
Comparison with \textbf{Transformer-based} Improvements (\%) & & & - & +2.27\%$^{***}$ & +0.00\%$^{}$ & +2.25\%$^{***}$ & +1.11\%$^{**}$ & +0.00\%$^{}$ \\
\hline
Comparison with \textbf{Graph-Transformer} Improvements (\%) & & & - & +0.00\%$^{}$ & +1.12\%$^{*}$ & +0.00\%$^{}$ & +0.00\%$^{}$ & +0.00\%$^{}$ \\
\bottomrule

\end{tabular}}}
\vspace{1mm}
\footnotesize\\
\textit{Significance.} $^{*}$p $<$ 0.05, $^{**}$p $<$ 0.01, $^{***}$p $<$ 0.001 indicate statistically significant improvements of CGMCL compared to the corresponding baseline models.
\label{table:Normal_Abnormal}
\end{table*}

\begin{table*}[htbp]
\centering
\caption{Comparison of the accuracy of multi-class classification performance between unimodal and proposed CGMCL on melanoma dataset. The best performance results are highlighted in \textbf{bold}, and "\_" indicates the second-best methods}
\resizebox{\linewidth}{!}{%
{\Huge
\begin{tabular}{@{}llcccccccccc@{}}
\toprule
\textbf{Model} & \textbf{Backbone} & \textbf{Image} & \textbf{Non-image} & \textbf{BWV} & \textbf{DaG} & \textbf{PIG} & \textbf{PN} & \textbf{RS} & \textbf{STR} & \textbf{VS} & \textbf{DIAG} \\ 
\midrule
\multicolumn{12}{l}{\textbf{Baseline}} \\
\hline
Logistic Model & - & \CheckmarkBold & \XSolidBrush & 0.83$\pm$0.00 & 0.58$\pm$0.20 & 0.51$\pm$0.15 & 0.68$\pm$0.10 & 0.77$\pm$0.00 & 0.73$\pm$0.10 & 0.83$\pm$0.03 & \underline{0.75$\pm$0.19} \\
Xgboost & - & \CheckmarkBold & \XSolidBrush & 0.82$\pm$0.00 & 0.49$\pm$0.14 & 0.68$\pm$0.12 & 0.67$\pm$0.09 &  0.76$\pm$0.00 & 0.74$\pm$0.10 & 0.84$\pm$0.03 & 0.74$\pm$0.19 \\
AdaBoost & - & \CheckmarkBold & \XSolidBrush & 0.83$\pm$0.03 & 0.54$\pm$0.09 & 0.64$\pm$0.04 & 0.60$\pm$0.03 & 0.77$\pm$0.02 & 0.68$\pm$0.04 & 0.84$\pm$0.01 & 0.66$\pm$0.03\\
\hline
\multicolumn{12}{l}{\textbf{Single Image (2-D)}} \\
\hline
ResNet18 & CNN & \CheckmarkBold  & \XSolidBrush  & 0.84$\pm$0.01 & 0.48$\pm$0.14 & 0.59$\pm$0.11 & 0.65$\pm$0.09 & 0.74$\pm$0.01 & 0.64$\pm$0.09 & 0.82$\pm$0.03 & 0.75$\pm$0.20 \\
2-layer  & CNN & \CheckmarkBold  & \XSolidBrush & 0.80$\pm$0.01 & 0.43$\pm$0.12 & 0.60$\pm$0.12 & 0.56$\pm$0.08 & 0.71$\pm$0.01 & 0.69$\pm$0.1 & 0.80$\pm$0.03 & 0.72$\pm$0.20 \\
\hline
\multicolumn{12}{l}{\textbf{Multimodal Fusion Methods}} \\
\hline
7-point\textsuperscript{\textdagger} \cite{kawahara2018seven}  & CNN & \CheckmarkBold  & \CheckmarkBold & 0.85 & 0.60 & 0.63 & 0.69 & 0.77 & 0.74 & 0.82 & 0.73 \\ 
HcCNN\textsuperscript{\textdagger}~\cite{bi2020multi} & CNN & \CheckmarkBold  & \CheckmarkBold  & 0.87 & 0.66 & 0.69 & 0.71 & 0.81  & 0.72 & 0.85 & 0.74  \\ 
AMFAM\textsuperscript{\textdagger} \cite{wang2022adversarial}  & GAN & \CheckmarkBold  & \CheckmarkBold  & 0.88 & 0.64 & 0.71 & 0.71 & 0.81 & 0.75 & 0.83 & 0.75 \\ 
FusionM4Net~\cite{tang2022fusionm4net}  & CNN & \CheckmarkBold  & \CheckmarkBold  & \textbf{0.89$\pm$0.00} & \underline{0.66$\pm$0.02} & 0.72$\pm$0.01 & 0.69$\pm$0.01 & \textbf{0.81$\pm$0.01}  & 0.76$\pm$0.01 & 0.82$\pm$0.01 &  0.76$\pm$0.01 \\ 
\hline

\multicolumn{12}{l}{\textbf{Vision-Transformer}} \\
\hline
SHViT~\cite{yun2024shvit}  & Vision-Transformer & \CheckmarkBold  & \CheckmarkBold  & 0.82±0.02 & 0.56±0.04 & 0.71±0.05 & 0.62±0.01 & 0.70±0.03 & 0.78±0.03 & 0.83±0.01 & 0.93±0.01 \\ 
MedFormer~\cite{xia2025medformer}  & Vision-Transformer & \CheckmarkBold  & \CheckmarkBold  & 0.81$\pm$0.02 & 0.59$\pm$0.01 & 0.69$\pm$0.03 & 0.59$\pm$0.03 & 0.72$\pm$0.01 & 0.72$\pm$0.30 & 0.83$\pm$0.25 & 0.92$\pm$0.00 \\ 
BIIGMA~\cite{roy2025background}  & Vector Sampling Module & \CheckmarkBold  & \CheckmarkBold  & 0.84$\pm$0.01 & 0.64$\pm$0.04 & 0.71$\pm$0.03 & 0.72$\pm$0.05 & 0.70$\pm$0.02 & 0.77$\pm$0.06 & 0.80$\pm$0.03 & 0.92$\pm$0.01 \\
PLRR~\cite{dong2024low}  & Vision-Transformer & \CheckmarkBold  & \CheckmarkBold  & 0.86$\pm$0.02 & 0.66$\pm$0.03 & 0.74$\pm$0.01 & 0.76 $\pm$0.05 & 0.77$\pm$0.01 & 0.78$\pm$0.03 & 0.87$\pm$0.01 & 0.94$\pm$0.01 \\



\hline
\multicolumn{12}{l}{\textbf{Proposed Model (2-D)}} \\
\hline
GCN+GCN & 2-layer CNN & \CheckmarkBold  & \CheckmarkBold & 0.81$\pm$0.01 & 0.60$\pm$0.06 & 0.73$\pm$0.08 & 0.74$\pm$0.20 & 0.73$\pm$0.01 & \textbf{0.80$\pm$0.04} & 0.81$\pm$0.03 & 0.92$\pm$0.02 \\
GAT+GAT & & \CheckmarkBold  & \CheckmarkBold & 0.79$\pm$0.02 & 0.46$\pm$0.13 & 0.60$\pm$0.12 & 0.69$\pm$0.20 & 0.71$\pm$0.03 & 0.69$\pm$0.09 & 0.80$\pm$0.03 & 0.93$\pm$0.01 \\
GCN+GCN & ResNet18 & \CheckmarkBold  & \CheckmarkBold &0.87$\pm$0.01 & 0.65$\pm$0.01 & \textbf{0.76$\pm$0.03} & \underline{0.75$\pm$0.05} & 0.78$\pm$0.02 & 0.76$\pm$0.02 & \underline{0.86$\pm$0.02} & \textbf{0.95$\pm$0.01} \\
GAT+GAT & & \CheckmarkBold  & \CheckmarkBold & \underline{0.87$\pm$0.01} & \textbf{0.67$\pm$0.03} & \underline{0.74$\pm$0.03} & \textbf{0.75$\pm$0.03} & \underline{0.78$\pm$0.02} & \underline{0.77$\pm$0.02} & \textbf{0.87$\pm$0.01} & 0.94$\pm$0.01 \\ 
\hline

Comparison with \textbf{Unimodal} Improvements (\%) & & & - & +3.57\%$^{***}$ & +39.58\%$^{***}$ & +26.67\%$^{***}$ & +15.38\%$^{***}$ & +5.41\%$^{***}$ & +15.94\%$^{***}$ & +6.10\%$^{***}$ & +26.67\%$^{***}$ \\
\hline
Comparison with \textbf{Multimodal Fusion} Improvements (\%) & & & - & +0.00\%$^{}$ & +1.52\%$^{}$ & +5.56\%$^{***}$ & +5.63\%$^{***}$ & +0.00\%$^{}$ & +5.26\%$^{***}$ & +2.35\%$^{***}$ & +25.00\%$^{***}$ \\
\hline
Comparison with \textbf{Vision-Transformer} Improvements (\%) & & & - & +1.16\%$^{**}$ & +1.52\%$^{}$ & +2.70\%$^{***}$ & +0.00\%$^{}$ & +1.30\%$^{**}$ & +2.56\%$^{*}$ & +0.00\%$^{}$ & +1.06\%$^{***}$ \\
\bottomrule

\end{tabular}%
}}
\raggedright
\vspace{1mm}
\footnotesize\\
\textit{Significance.} $^{*}$p $<$ 0.05, $^{**}$p $<$ 0.01, $^{***}$p $<$ 0.001 indicate statistically significant improvements of CGMCL compared to the corresponding baseline models. \textsuperscript{\textdagger}Denotes the average of accuracy.
\label{table:melanoma_table}
\end{table*}

\begin{figure*}
\centering
\includegraphics[width=1\textwidth]{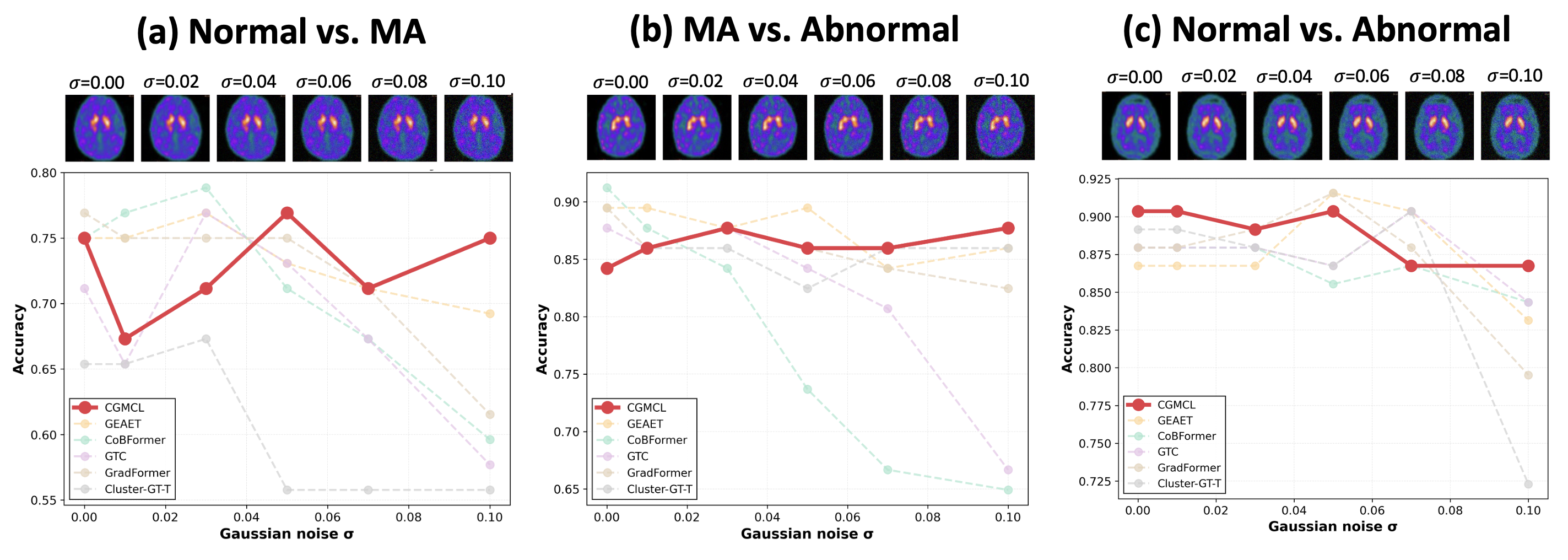}
\caption{Robustness evaluation of the proposed CGMCL model under Gaussian noise perturbations ($\sigma=0.01$–$0.10$) for three SPECT-based PD subtyping tasks.}
\label{fig:noise_ACC}
\end{figure*}

\section{EXPERIMENTS}

\subsection{Experiments Setting}

In this work, the experiments were conducted using the PyTorch framework on a workstation equipped with an NVIDIA GeForce RTX 3070 GPU (8 GB VRAM) with CUDA 12.2 support. The models were jointly optimized with the Adam optimizer (learning rate = 0.001) for 300 epochs and a batch size of 32. As shown in Table~\ref{tab:model_performance}, CGMCL achieves the best balance between accuracy and computational efficiency. It has fewer parameters (15.07 M) and lower latency (8.879 ms) than other graph-transformer models, with comparable GFLOPs (68.052). Its training time (32.94 s) is shorter than most transformer-based baselines, demonstrating faster convergence. Although GPU usage (1997.5 MB) is slightly higher than GradFormer, it remains efficient relative to GTC and GEAET. Overall, CGMCL achieves high performance with low computational cost, indicating that the proposed cross-graph contrastive learning framework offers a good trade-off between accuracy and efficiency for multimodal medical imaging tasks.

\subsection{Baseline Methods Comparison}

To fairly compare the effectiveness of CGMCL across various methods for both the PD and melanoma datasets, we conducted a quantitative analysis of baseline method comparisons. The results for PD, segmented into three subtypes, are presented in \textcolor{blue}{Table} \ref{table:Normal_MA} through \textcolor{blue}{Table} \ref{table:Normal_Abnormal}, and for the melanoma multi-class dataset in \textcolor{blue}{Table} \ref{table:melanoma_table}.  Initially, we employed conventional machine learning algorithms to assess the impact of a single modality on meta-features. This analysis applied logistic regression, XGBoost, random forest, support vector machines (SVM)~\cite{Hsu_2019}, and AdaBoost~\cite{freund1996experiments} to both datasets.   To further evaluate the performance of our proposed CGMCL model in PD classification, we compared it to unimodal two-layer CNN models.  We then extended the analysis using SPECT images with three slices to compare different 3D-CNN architectures. When comparing machine learning methods that utilize only meta-features against CNN-based unimodal approaches, nonlinear logistic and tree-based methods (i.e., XGBoost and AdaBoost) achieved accuracies between 0.58 and 0.63 for distinguishing Normal from MA in PD. This indicates that relying solely on quantitative data may result in losing other informative features, such as those derived from images. In contrast, unimodal image feature extraction models substantially improved classification accuracy for the three PD subtypes. Specifically, ResNet18 increased accuracy by 16\% for Normal vs. MA, 12.5\% for Normal vs. Abnormal, and 1.16\% for Normal vs. Abnormal compared to logistic regression. Conversely, for the melanoma dataset, meta-features in ML methods outperform in the DaG, PN, and VS categories.

\subsection{Compare with the SOTA Methods}

In this work, we examined the effectiveness of our proposed CGMCL framework by conducting extensive comparisons with a range of state-of-the-art (SOTA) unimodal and multimodal models, such as MHCA, DeAF, and TriFormer. These models employ 3D tensor-based CNN architectures as feature extractors jointly trained with clinical meta-features, transformer-based multimodal fusion methods, and recent graph-transformer hybrid models. The quantitative results are presented for both the Parkinson's disease dataset (Tables~\ref{table:Normal_MA}, \ref{table:MA_Abnormal}, \ref{table:Normal_Abnormal}) and the melanoma dataset (Table~\ref{table:melanoma_table}). Our experimental results yielded the following findings:



\begin{enumerate}
    \item Compared to unimodal CNN-based feature extractors (i.e., 2-layer CNN, 3D-CNN, VGG19, and ResNet18), the three SOTA models using multimodal 3D-CNNs did not demonstrate substantially higher ACC across the three PD subtypes. However, individual metrics improved for certain models. Specifically, DeAF showed enhancements in specificity (SPE) and positive predictive value (PPV) for Normal vs. MA  (SPE = 0.76) and for MA vs. Abnormal (SPE = 0.91, PPV = 0.93). MHCA also improved sensitivity (SEN) for Normal vs. Abnormal (SEN = 0.92).
    \item We further evaluated the cross-modal graph fusion capability of our proposed CGMCL model in classifying the three PD subtypes. From \textcolor{blue}{Table} \ref{table:Normal_MA}, using GAT as the cross-modal fusion method with ResNet18 achieves metric values between 0.73 and 0.76, comparable to the SOTA model using 3D-CNNs or unimodal approaches (ACC $\uparrow$ 4.1\%, PPV $\uparrow$ 3.9\%). This indicates that the fusion of clinical and imaging data can more effectively assist in identifying difficult-to-diagnose early PD symptoms. Moreover, there is a noticeable improvement in classification performance for the other two subtypes (e.g., MA vs. Abnormal and Normal vs. Abnormal).

    \item When comparing our proposed CGMCL with other SOTA models on the melanoma dataset, although CGMCL’s two-class classification performance for BWV and RS was lower than that of the latest FusionM4Net model, it achieved higher accuracy in most three-class classifications (DaG, PIG, PN, VS, and DIAG). This superiority in multi-class tasks is due to the model’s ability to handle imbalanced sample issues and fine-grained categories across multiple classes. Additionally, our CGMCL model demonstrated consistent performance across multi-class challenges. Our approach achieved an impressive ACC of 0.95 for the five-class DIAG task.

    \item We compared CGMCL with recent graph-transformer architectures (GEAET, CoBFormer, GTC, GradFormer, Cluster-GT) that exhibit strong capability in learning structural dependencies among patients and vision-transformer models (SHViT, MedFormer, BIIGMA, PLRR). As shown in Tables~\ref{table:MA_Abnormal} to \ref{table:melanoma_table}, CGMCL achieves competitive performance across PD subtypes, with notable improvements in Normal vs. MA classification (ACC=0.76, +2.70\% over best graph-transformer baseline). For melanoma (Table~\ref{table:melanoma_table}), CGMCL demonstrates superior multi-class classification in five categories (DaG, PIG, PN, VS, DIAG), achieving 95\% ACC in the five-class DIAG task versus PLRR (94\%) and BIIGMA (92\%).
    
\end{enumerate}


\subsection{Robustness Analysis}

Unlike the popular high-quality publicly available PPMI dataset for PD analysis, prior research on PPMI data has shown that high-quality data often leads to improved model accuracy and stability~\cite{ding2021dopamine}. However, there is a lack of more refined early-stage PD diagnosis (\textit{e.g.}, MA) and realistic noisy medical images commonly found in most hospitals. Therefore, to simulate common image degradation scenarios in clinical practice, we added varying levels of Gaussian noise ($\sigma=0.00$-$0.10$) to the input images to analyze the classification stability of CGMCL, as shown in Fig.~\ref{fig:noise_ACC}. Compared to baseline models (GEAET, CoBFormer, GTC, GradFormer, Cluster-GT-T), which exhibited significant performance degradation under the same noise conditions with accuracy drops of 10--20\%, CGMCL demonstrated superior robustness. When noise intensity increased from $\sigma=0.00$ to $\sigma=0.10$, CGMCL's accuracy only fluctuated slightly from 0.75 to the 0.67--0.77 range in the Normal vs.\ MA task, maintained a high level of 0.84--0.88 in the MA vs.\ Abnormal task, and remained stable at 0.87--0.90 in the Normal vs.\ Abnormal task. This robustness stems from the cross-graph contrastive learning and inter-modality feature scaling, which jointly enhance noise tolerance and feature consistency.

\section{Ablation Study}

To evaluate the contribution of each component to CGMCL, we conduct ablation studies examining the effectiveness of different components on overall classification performance. Specifically, we analyze: (1) the alignment of multimodal representations (Section~\ref{sec:alignment}), (2) the sensitivity of the parameter $K$ for $K$-nearest neighbors used in constructing the modality graph (Section~\ref{sec:K_NN_graph}), and (3) the IMFES module and the influence of the balancing weight coefficient $\beta$ between supervised classification loss and cross-graph contrastive loss (Sections~\ref{sec:IMFES_module} and \ref{sec:beta}).


\subsection{Multimodal Features Alignment in KL Divergence Converge}\label{sec:alignment}

\begin{figure}
\centering
\includegraphics[width=0.5\textwidth]{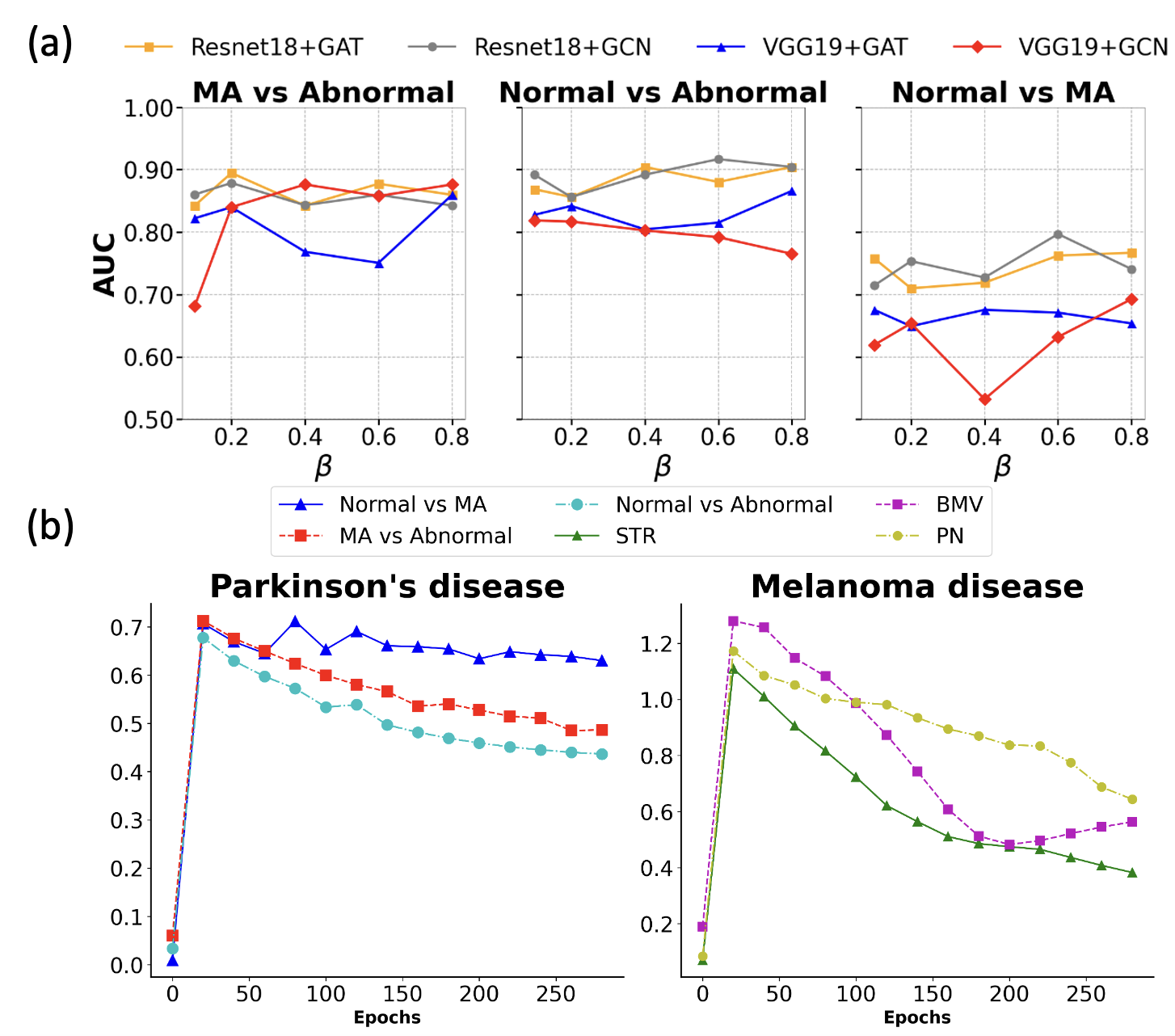}
\caption{The ablation study of the weighted loss coefficient $\beta$ is shown in Panel (a), while Panel (b) illustrates the convergence of the two modality features' alignment using KL divergence.}
\label{fig:beta_KL}
\end{figure}

The primary challenge in multimodal disease classification is ensuring proper alignment of features across different modalities during the fusion process. We evaluated the two learned representations $\hat{Z}^{I}$ and $\hat{Z}^{C}$ by converting them into probability matrices using the sigmoid function $\sigma(\cdot)$ and then calculating their Kullback-Leibler (KL) divergence, expressed as $\mathrm{KL}(\sigma(\hat{Z}^{I}) \mathbin\Vert \sigma(\hat{Z}^{C}))$. The results are illustrated in \textcolor{blue}{Fig.} \ref{fig:beta_KL} (b). A key observation is that the KL divergence across the three PD subtypes shows remarkable fluctuations after the first 50 epochs. This is especially apparent when comparing the Normal vs. MA subtypes, which require additional epochs to reach convergence. Melanoma showed clear convergence in terms of KL divergence. At the same time, PN and STR experienced a sharper decline in KL divergence during the first 50 epochs, indicating that the model can quickly differentiate these pathological markers. After 100 epochs, BWV exhibited more substantial changes in divergence compared to PN and STR. However, PN and STR demonstrated more stable convergence overall. PD maintained relatively low and stable KL divergence values, ranging from 0.4 to 0.7. In contrast, Melanoma initially had higher KL divergence values (near 1.2) but rapidly decreased and eventually stabilized between 0.4 and 0.6. These findings highlight how the multimodal model effectively captures distinct distributions among categories in these two domains and how CGMCL enhances overall training convergence.

\begin{figure}
\centering
\includegraphics[width=0.5\textwidth]{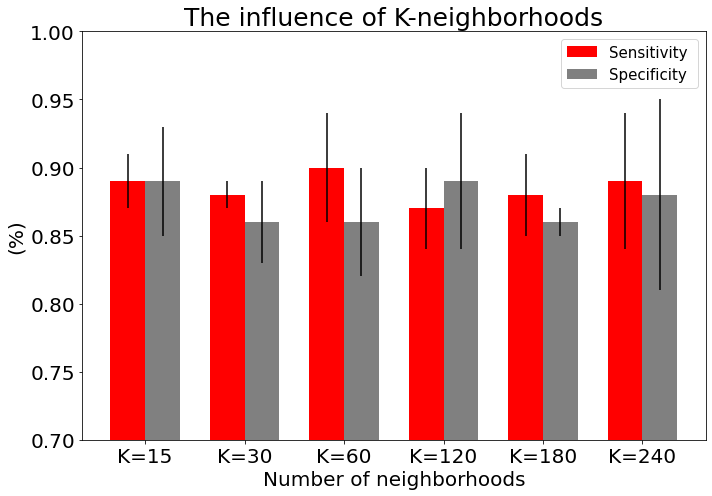}
\caption{The mean and standard deviation of our proposed CGMCL's performance in terms of sensitivity and specificity for PD, based on varying numbers of K-neighborhoods for constructing the modality graph.}
\label{fig:KNN_ablation}
\end{figure}

\subsection{Impact of K-Neighborhood Graph Construction}\label{sec:K_NN_graph}

Furthermore, to examine the robustness of our CGMCL model under different graph construction settings, we analyzed the influence of varying the number of $K$-neighborhoods in building the modality graph. As shown in Fig.~\ref{fig:KNN_ablation}, the mean and standard deviation of sensitivity and specificity were computed across different $K$ values for PD classification. The results indicate that the model maintains stable performance within a moderate range of $K$ (from 60 to 120), suggesting that CGMCL is relatively insensitive to small fluctuations in neighborhood size and effectively captures patient similarity relationships across modalities.

\subsection{IMFES Module Ablation}~\label{sec:IMFES_module}

To validate CGMCL in module ablation for multimodal classification performance in latent space, we compare (1) \textbf{w/o Concatenate}, which excludes the concatenation from \textcolor{blue}{Eqs.}~\ref{eq:eq9}, and ~\ref{eq:eq10}, (2) \textbf{w/o IMFES}, which removes the IMFES module from \textcolor{blue}{Eqs.}~\ref{eq:eq11}, and ~\ref{eq:eq12} under different backbones with the entire module. As shown in \textcolor{blue}{Table} \ref{table:IMFES_ablation}, the experimental results indicate that the GAT+GAT combination with a ResNet18 backbone yields a slight AUC increase for the more distinct Normal vs. Abnormal subtypes, compared to models without concatenation or IMFES. Notably, CGMCL demonstrates a 5.5\% improvement in the Normal vs. MA classification. Furthermore, it can inferred that utilizing various multi-level concatenation layers and weighted inter-modality distributions effectively enhances the extraction of critical feature information.

\begin{table}
\centering
\small
\caption{Evaluating classification performance and ablation study of module components in CGMCL}
\setlength{\tabcolsep}{0.5pt} 
\begin{tabular}{|c|c|c|c|c|}
\hline
\textbf{Model} & \textbf{Backbone} & \textbf{w/o Concatenate} & \textbf{w/o IMFES} & \textbf{CGMCL} \\
\hline
\multicolumn{5}{|c|}{\textbf{Parkinson (Normal vs MA)}} \\
\hline
GCN+GCN & 2-layer CNN & $0.58 \pm 0.07$ & $0.53 \pm 0.05$ & \textbf{0.66 $\pm$ 0.03 }\\
GAT+GAT & 2-layer CNN & $0.62 \pm 0.06$ & $0.56 \pm 0.08$ & \textbf{0.66 $\pm$ 0.04} \\
GCN+GCN & ResNet18 & $0.69 \pm 0.04$ & $0.71 \pm 0.03$ & \textbf{0.72 $\pm$ 0.05} \\
GAT+GAT & ResNet18 & $0.73 \pm 0.04$ & $0.73 \pm 0.04$ & \textbf{0.77 $\pm$ 0.02} \\
\hline

\multicolumn{5}{|c|}{\textbf{Parkinson (MA vs. Abnormal)}} \\
\hline
GCN+GCN & 2-layer CNN & $0.76 \pm 0.12$ & $0.63 \pm 0.16$ & \textbf{0.82 $ \pm $ 0.01} \\
GAT+GAT & 2-layer CNN & $0.80 \pm 0.04$ & $0.71 \pm 0.13$ & \textbf{0.85 $ \pm $ 0.02} \\
GCN+GCN & ResNet18 & $0.74 \pm 0.13$ & $0.82 \pm 0.05$ & \textbf{0.86 $ \pm $ 0.02} \\
GAT+GAT & ResNet18 & \textbf{0.88 $ \pm $ 0.02} & $0.83 \pm 0.03$ & 0.85 $ \pm $ 0.03 \\
\hline
\multicolumn{5}{|c|}{\textbf{Parkinson (Normal vs. Abnormal)}} \\
\hline
GCN+GCN & 2-layer CNN & $0.84 \pm 0.03$ & $0.84 \pm 0.02$ & \textbf{0.86 $ \pm $ 0.01} \\
GAT+GAT & 2-layer CNN & $0.86 \pm 0.02$ & $0.77 \pm 0.11$ & \textbf{0.86 $ \pm $ 0.02} \\
GCN+GCN & ResNet18 & $0.79 \pm 0.11$ & $0.88 \pm 0.02$ & \textbf{0.89 $ \pm $ 0.01}\\
GAT+GAT & ResNet18 & $0.88 \pm 0.02$ & $0.87 \pm 0.01$ &  \textbf{0.89 $ \pm $ 0.01}  \\
\hline

\end{tabular}
\label{table:IMFES_ablation}
\end{table}

\begin{figure}
\centering
\includegraphics[width=0.5\textwidth]{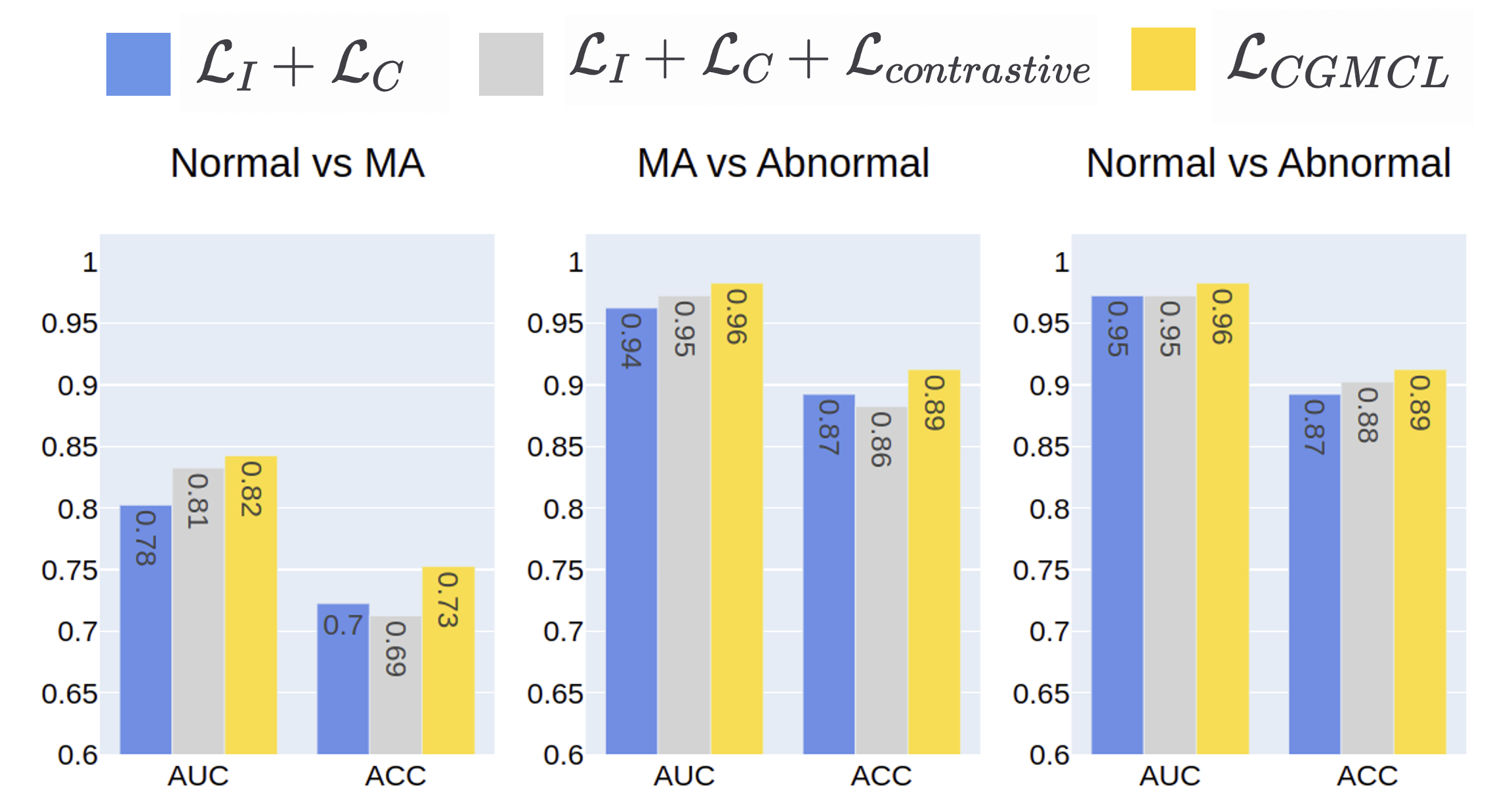}
\caption{The AUC and ACC performance of ablation experiments on various loss function combinations across the three PD subtypes.}
\label{fig:loss_ablation}
\end{figure}

\begin{figure*}
\centering
\includegraphics[width=1\textwidth]{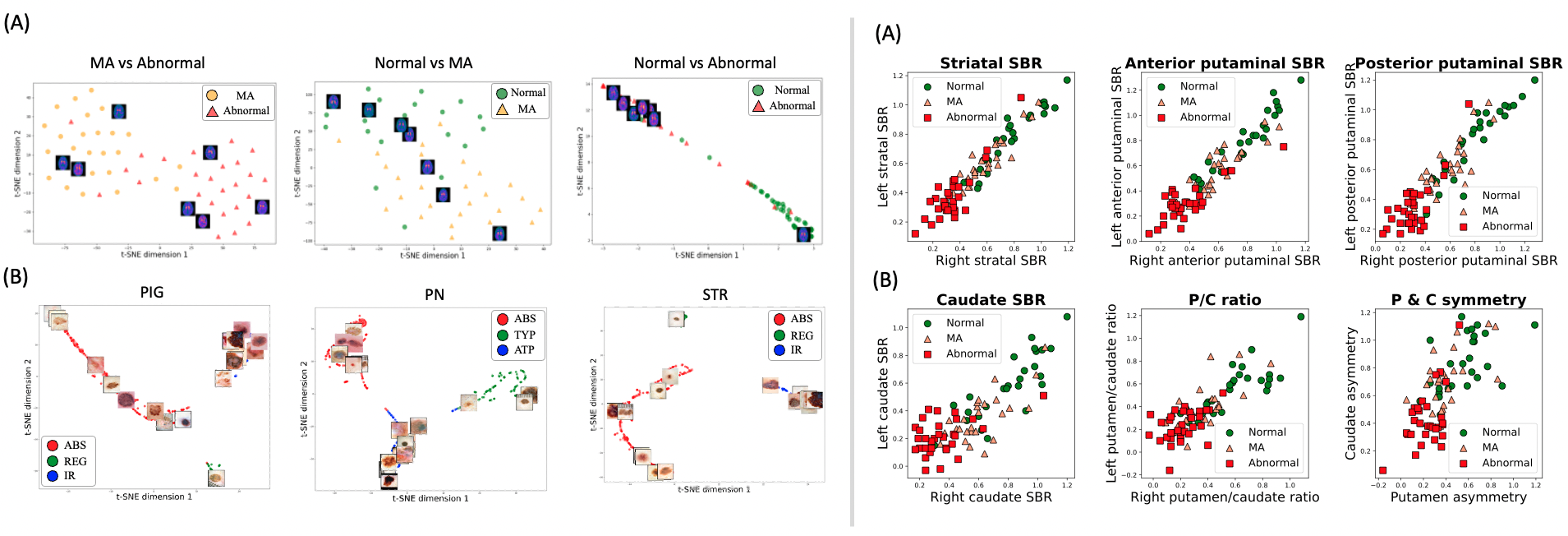}
\caption{Visualization of CGMCL multimodal representations and clinical parameter analysis. (Left) Visualization of CGMCL predictions incorporating twelve indicators from twelve DaTQUANT parameters. (Right) t-SNE visualization of CGMCL low-dimensional embeddings on two multimodal datasets.}
\label{fig:vis}
\end{figure*}

\begin{figure*}
\centering
\includegraphics[width=1\textwidth]{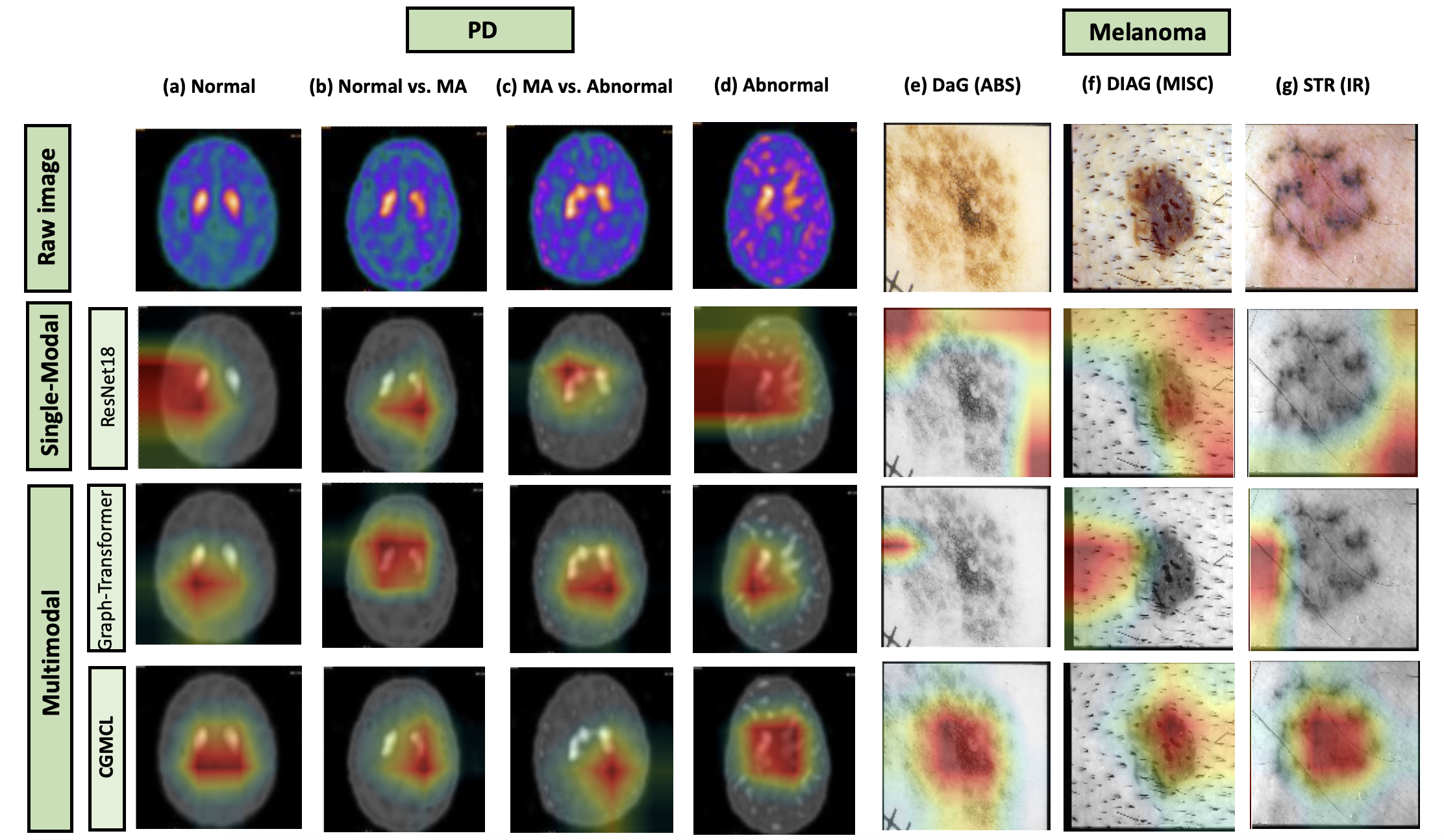}
\caption{Grad-CAM visualization of multimodal feature maps across disease domains. Compared with unimodal and graph-transformer models, CGMCL produces more localized and clinically relevant Grad-CAM activations within striatal and lesion regions, indicating enhanced interpretability and cross-modal feature alignment.}
\label{fig:grad-cam}
\end{figure*}

\subsection{Objective Function Ablation}\label{sec:overall_loss}

The well-structured contrastive term within the cross-graph modal objective function plays a pivotal role in aligning multimodal feature distributions and enhancing fusion consistency during model optimization. We evaluate the combination of the supervised classification losses and the cross-graph contrastive loss with balancing parameter $\beta$ for CGMCL's multimodal fusion performance.

\paragraph{(a) \textbf{Effect of Loss Components}}

We first compared several configurations of the objective function to examine the effect of the contrastive and structural consistency losses. As illustrated in ~\textcolor{blue}{Fig.} \ref{fig:loss_ablation}, incorporating the contrastive loss significantly improved the alignment of multimodal representations. The results demonstrate that the full $\mathcal{L}_{CGMCL}$ yields the highest fusion performance. Specifically, for the MA vs. Abnormal classification, $\mathcal{L}_{CGMCL}$ surpasses $\mathcal{L}_{I} + \mathcal{L}_{C}$ by 2.1\% in AUC and 2.2\% in ACC, while for the Normal vs. Abnormal task, improvements of 1.1\% are observed in both metrics. Notably, in the most challenging Normal vs. MA classification, which is critical for early Parkinson's detection, the full weighted loss achieves a 5.1\% AUC and 4.3\% ACC gain over the next best unweighted setting. These improvements indicate that the contrastive component enhances the CGMCL’s ability to capture subtle multimodal correlations and latent disease patterns in the shared embedding space.

\paragraph{\textbf{(b) Sensitivity of the Weighting Parameter $\beta$}}\label{sec:beta}



We further investigated how varying $\beta$ affects the interplay between supervised and contrastive objectives. As illustrated in \textcolor{blue}{Fig.} \ref{fig:beta_KL}, models employing different CNN backbones (e.g., ResNet18, VGG19) in conjunction with GCN or GAT graph modules exhibit stable performance across a wide $\beta$ range. In the MA vs. Abnormal and Normal vs. Abnormal tasks, ResNet18 coupled with either GAT or GCN achieves an AUC close to 0.90 across  $\beta \in [0.4, 0.8]$, indicating robustness to weighting variations. For the more difficult Normal vs. MA task, optimal performance (AUC = 0.80) occurs at $\beta = 0.60$, suggesting that moderate weighting of the contrastive term yields the best balance between modality discrimination and shared-space alignment. These ablation studies validate the necessity of an adaptive $\beta$ setting to ensure optimal cross-modal optimization and highlight the role of $\beta$ in maintaining equilibrium between intra-modal classification and inter-modal consistency.



\section{Visualization of Multimodal Representation}
\subsection{Identification of PD subtype trajectories}


As illustrated in \textcolor{blue}{Fig.} \ref{fig:vis} (Left), the embedding trajectory presents a smooth transition from Normal to MA and Abnormal, reflecting the clinical progression of dopaminergic degeneration. Notably, the CGMCL performance on the Normal versus MA distinction becomes a critical boundary for early disease prevention, suggesting improved sensitivity to subtle early-stage abnormalities. This enhanced separability indicates that the integration of imaging and quantitative DaTQUANT features effectively reduces representation ambiguity and supports early PD diagnosis. Furthermore, the final-layer representations of each patient, derived from the integrated image and meta-features, were embedded into a two-dimensional latent space as shown in \textcolor{blue}{Fig.} \ref{fig:vis} (Right). A particularly meaningful observation lies in the boundary region between Normal and MA clusters, which represents early-stage PD patients with subtle dopaminergic degeneration. In this transitional zone, the embeddings are less dispersed compared to unimodal baselines, suggesting that CGMCL successfully reduces representation ambiguity by leveraging complementary information from both SPECT images and DaTQUANT-derived quantitative metrics.

\begin{figure}
\centering
\includegraphics[width=0.5\textwidth]{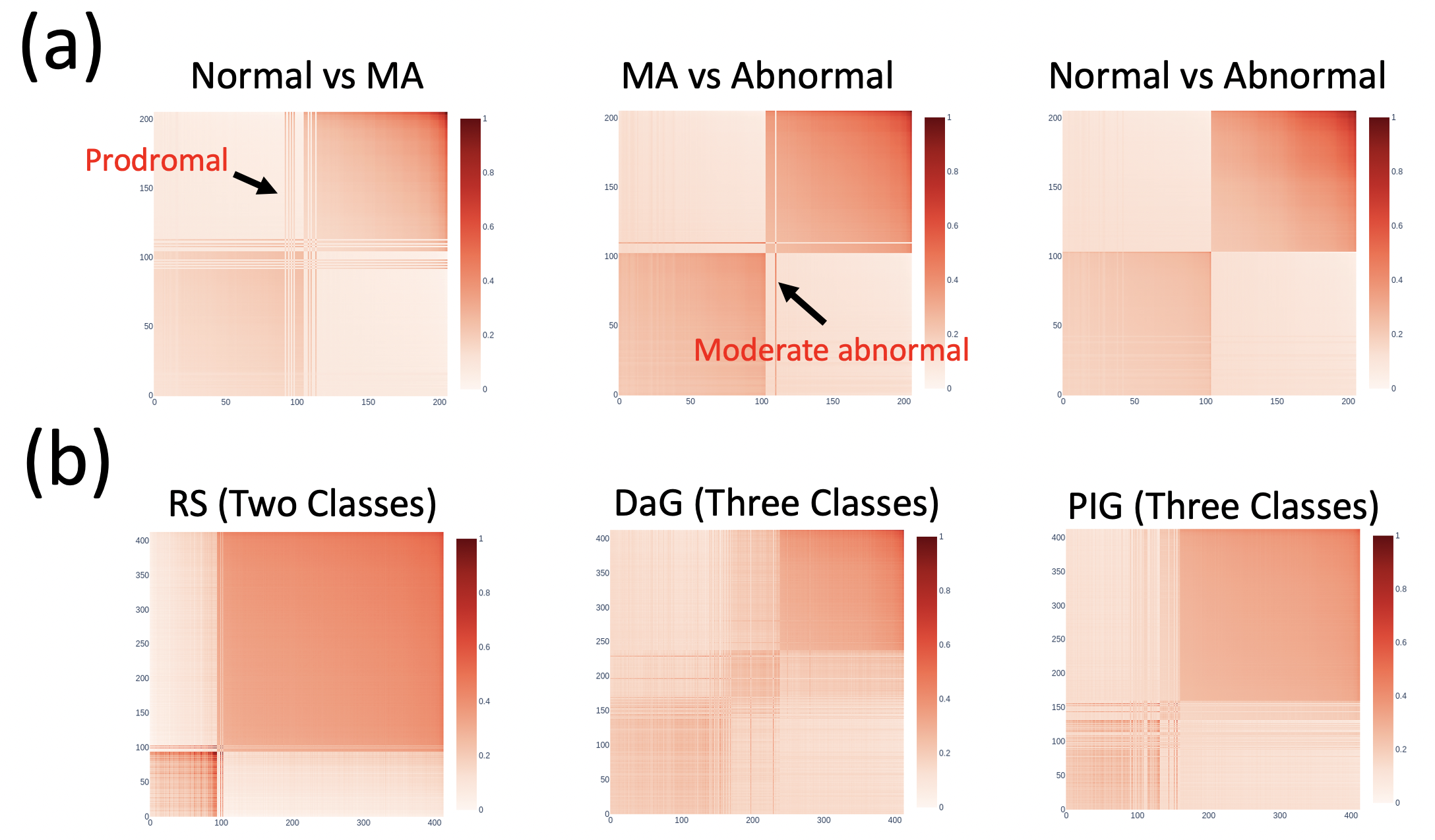}
\caption{Visualization of common similarity matrices for modality clustering results between patients $i$ and $j$: Panel (a) shows the intrinsic similarity matrices for three PD patients. At the same time, panel (b) displays the similarity matrices of CGMCL clustering for different imbalanced classes in melanoma.}
\label{fig:clustering_results}
\end{figure}


\begin{figure*}
    \centering
    \includegraphics[width=1\textwidth]{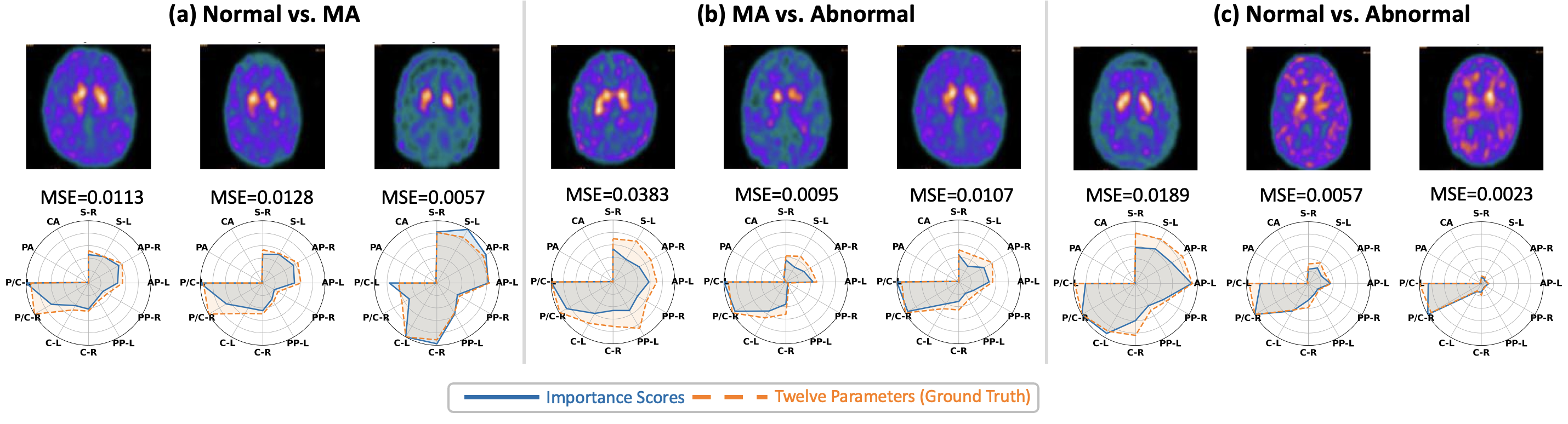}
    \caption{The radar chart of explainable modality importance scores for quantitative analysis of overall striatal (R-L) indicators across different PD subtypes.}
    \label{fig:radar}
\end{figure*}

\subsection{Multi-Class Trajectories in Melanoma Embedding}

For multi-class melanoma classification, the seven-point criteria categorize skin lesion characteristics into groups based on similarity, with physicians assigning corresponding scores. \textcolor{blue}{Fig.} \ref{fig:vis} (B) illustrates how CGMCL generates low-dimensional representations based on similar features for PIG, PN, and STR, which correspond to feature labels (ABS, REG, IR), (ABS, TYP, ATP), and (ABS, REG, IR), respectively. In \textcolor{blue}{Fig.} \ref{fig:vis}, for PN, the feature score of ATP is 2 (e.g., atypical pigment network, blue-whitish veil, and irregular vascular structures). It is evident that ATP has a more distinct and severe scoring, showing clear separation from ABS and TYP. Some ATP positions are located within TYP, indicating that certain TYP cases may progress toward more malignant features. Compared to low-dimensional representations in previous studies, our CGMCL model integrates more precise multimodal information into feature representations and provides diagnostic insights, addressing the lack of interpretability often associated with meta-features in existing research work.


\subsection{Heatmaps Visualization of Multimodal Feature Activations}

To further evaluate the interpretability and feature localization capability of the proposed CGMCL framework across different disease domains. We use the Grad-CAM to visualize activation maps across disease domains, as shown in Fig.~\ref{fig:grad-cam}.  For PD classification, CGMCL produces highly localized activations within the striatum, particularly in the caudate and putamen regions where dopaminergic degeneration occurs~\cite{lopez2023dopaminergic,nazari2022explainable,saeed2020neuroimaging}.  In contrast, unimodal ResNet18 exhibits diffuse activations beyond striatal regions, while graph-transformer models show improved but inconsistent localization with scattered attention across multiple brain regions. Notably, in the normal vs. MA classification, CGMCL generates focused activations in the posterior putamen, consistent with early PD pathology. We can observe in melanoma dataset, the CGMCL significance captures lesion-specific morphological features, including concentrated activations on irregular pigmented structures for DaG, discriminative patterns for multi-class DIAG, and radial streaming patterns for STR. Fig.~\ref{fig:grad-cam} shows that CGMCL's unified multi-modality fusion yields more focused attention on disease-concentrated regions compared to other approaches.

\section{Discussion}

Multimodal medical image fusion faces challenges from heterogeneous distributions, prevalent noise, and the need for effective fusion methods for accurate diagnosis. Our proposed CGMCL framework integrates multimodal medical data through cross-graph modal fusion~\cite{azam2022review,kalamkar2023multimodal}. Specifically, CGMCL enables (1) cross-modal patient similarity clustering, and (2) quantitative meta-feature importance estimation for clinical decision support, and (3) anatomically interpretable attention visualization that aligns with established disease progression patterns. 

As shown in Fig.~\ref{fig:clustering_results}, the learned similarity matrices capture distinct clustering patterns that quantify patient relationships in the latent space. For PD subtypes (Fig.~\ref{fig:clustering_results} (a)), Normal vs. Abnormal displays strong diagonal concentration (similarity > 0.8), indicating well-separated embeddings, while Normal vs. MA shows more diffused patterns (0.4–0.6), consistent with the 76\% vs. 90\% accuracy gap in Tables~\ref{table:Normal_MA} to \ref{table:Normal_Abnormal} and highlighting the challenge of early-stage detection. The MA vs. Abnormal matrix exhibits intermediate similarity (0.7), suggesting gradual feature separation with disease progression. For melanoma (Fig.~\ref{fig:clustering_results} (b)), despite class imbalance (RS and DaG having ~30\% fewer samples), CGMCL maintains concentrated diagonal patterns (> 0.75), demonstrating effective neighborhood aggregation and robustness across multimodal contexts.

To identify the neurodegenerative regions from PD, the Grad-CAM visualizations (Fig.~\ref{fig:grad-cam}) under four conditions (Normal, Normal vs. MA, MA vs. Abnormal, and Abnormal) consistently highlight the striatum, especially the posterior putamen~\cite{drori2025multiparametric,ortiz2019parkinson}. As disease severity increases, the activation extends anteriorly toward the caudate nucleus. This spatial pattern is consistent with established DAT imaging findings in PD, where dopaminergic loss begins in the posterior putamen and progresses anteriorly, typically showing lateral asymmetry~\cite{lopez2023dopaminergic,calle2019identification,ogawa2018role,fiorenzato2021asymmetric}. 

Furthermore, CGMCL provides an effective visualization tool for meta-feature importance to support clinical decision-making, as shown in Fig.~\ref{fig:radar}. For Normal vs. MA (Fig.~\ref{fig:radar} (a)), posterior putamen parameters (PP-R, PP-L) show highest importance (0.85-0.90), exceeding anterior putamen (0.60-0.65) and caudate (0.45-0.50), consistent with Lewy body pathology progressing from posterior to anterior regions~\cite{lee2023substantia}. For MA vs. Abnormal (Fig.~\ref{fig:radar} (b)), importance becomes balanced across regions (0.75-0.88), with elevated P/C ratio (0.82-0.85) and caudate parameters (0.80-0.83), reflecting widespread bilateral degeneration in advanced disease~\cite{shin2007use,ciesielska2017depletion}. For Normal vs. Abnormal (Fig.~\ref{fig:radar} (c)), all parameters show uniformly high importance (0.82-0.92), with striatal SBR reaching peak scores (0.91-0.92). This progression pattern demonstrates posterior putamen dominance in early detection, expanding to bilateral putamen-caudate involvement, and then global striatal changes~\cite{pitcher2012reduced}.

These experimental findings collectively confirm the effectiveness of our cross-graph modal fusion framework, which preserves intra-modal patient structures while achieving cross-modal alignment through contrastive learning, reducing KL divergence from 1.2 to 0.4--0.6 (Fig.~\ref{fig:beta_KL} (b)). The IMFES module adaptively balances modality contributions, improving performance by 5.5\% over simple concatenation (Table~\ref{table:IMFES_ablation}), while graph attention jointly models imaging and clinical similarities, enhancing reliability in the challenging Normal to MA gray zone. The Grad-CAM visualizations in Fig.~\ref{fig:grad-cam} reveal that CGMCL yields more clinically meaningful attention patterns compared to single-modal baselines in the PD dataset. Specifically, the model's attention shifts from bilaterally symmetric putamen in healthy controls to asymmetrically distributed striatal regions in PD patients, consistent with the characteristic dopaminergic degeneration patterns illustrated in Fig.~\ref{fig:radar}. 



\section{Limitations and Future Work}

Although the proposed CGMCL framework demonstrates promising performance in multimodal PD classification, certain limitations remain. In early-stage PD, patients typically exhibit asymmetric degeneration of the striatum, where one side tends to deteriorate first, but the affected side varies among individuals. Eventually, both sides decline to a similar level.

\begin{figure}
\centering
\includegraphics[width=0.5\textwidth]{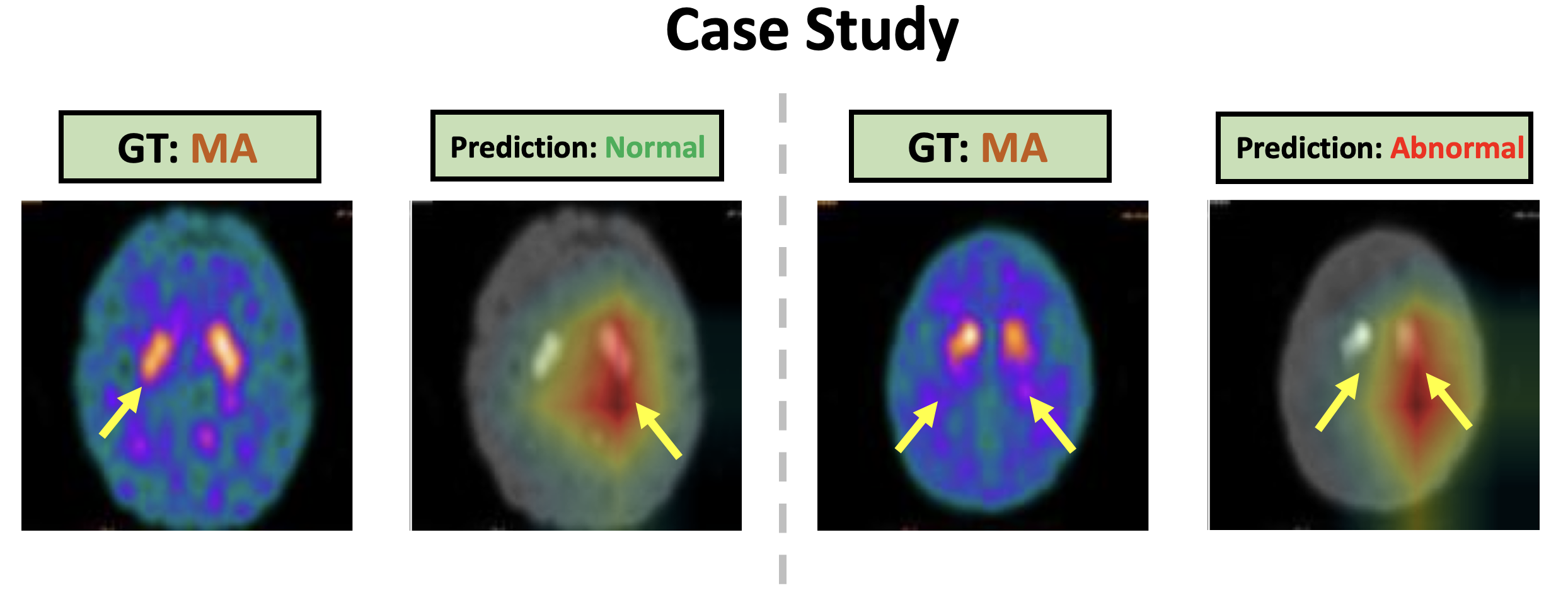}
\caption{Case study on failure analysis in mildly abnormal PD classification with the proposed CGMCL model}
\label{fig:case}
\end{figure}

The case study results in Fig.~\ref{fig:case} reveal several limitations that warrant further discussion. CGMCL performs inconsistently on MA cases representing early-stage PD. As shown in the left case of Fig.~\ref{fig:case}, the model misclassified an MA scan as Normal (false negative). The corresponding attention map indicates that the network primarily focused on high-intensity regions of the striatum while overlooking the subtle reduction in the putamen signal. Conversely, the right case illustrates the opposite error (over-sensitivity). Although the ground truth label was MA, the model predicted Abnormal (false positive), overemphasizing symmetric bilateral signal decreases. These findings suggest that the image modality tends to dominate during multimodal fusion, leading to an overreliance on pixel-level variations.

In future work, we will address these issues by (1) incorporating region-specific graph nodes (e.g., putamen and caudate ROIs) to enhance spatial representation, and (2) developing a larger multimodal framework to improve fine-grained ROI detection for more precise classification. Alternatively, we will explore CLIP-based multimodal models~\cite{radford2021learning,zhang2023biomedclip} with clinical text data to further enhance interpretability and diagnosis.

\section{Conclusion}
This study introduces a novel multimodal fusion framework that effectively integrates medical imaging and clinical data through dual-graph modeling and contrastive learning within a unified latent space. Our approach effectively combines medical imaging data with clinical features, enhancing multimodal fusion. By aligning heterogeneous modalities in a unified latent space, CGMCL enhances feature representation, interpretability, and robustness in the classification of two multimodal datasets. Additionally, the IMFES module further mitigates modality imbalance, enabling the model to adaptively balance contributions from diverse data sources. These results show that CGMCL has strong potential for broader applications in diverse multimodal disease studies and future clinical decision-support systems.


\section*{CRediT authorship contribution statement}
\textbf{Jun-En Ding:} Conceptualization, Methodology, Software, Writing – original draft, Writing – review \& editing. 
\textbf{Chien-Chin Hsu:} Data collection and medical guidance. 
\textbf{Chi-Hsiang Chu:} Statistical guidance. 
\textbf{Shuqiang Wang:} Methodological guidance. 
\textbf{Feng Liu:} Supervision, Methodology, Validation.

\subsection{Declarations}

\textbf{Ethics approval and consent to participate} \\
This study was approved by the Institutional Review Board of Taipei Veterans General Hospital, Taipei, Taiwan. Given the retrospective nature of the study, the requirement for informed consent was waived by the IRB.

\section{Acknowledgment:} We are grateful to the Department of Nuclear Medicine at Kaohsiung Chang Gung Memorial Hospital for providing us with comprehensive data and data labeling support.  

\subsection{Funding}
This study received no funding support.

\bibliography{sample}

\end{document}